\RequirePackage{fixltx2e}
\documentclass[aps,cha,twocolumn,superscriptaddress,longbibliography]{revtex4-1}

\usepackage{multirow}

\pdfoutput=1
\usepackage{url}
\usepackage{amsmath}
\usepackage{amsfonts}
\usepackage{calc}
\usepackage{graphicx}
\long\def\/*#1*/{}
\usepackage [english]{babel}
\usepackage [autostyle, english = american]{csquotes}
\usepackage{color}
\MakeOuterQuote{"}
\usepackage{tabularx}
\usepackage{graphicx}

\begin{document}
\title{Hybrid Monte Carlo with Chaotic Mixing}
\author{Nirag Kadakia} 
\affiliation{University of California at San Diego, Department of Physics, 9500 Gilman Drive, San Diego, CA 92093 USA}

\begin{abstract}

We propose a hybrid Monte Carlo (HMC) technique applicable to high-dimensional multivariate normal distributions that effectively samples along chaotic trajectories. The method is predicated on the freedom of choice of the HMC momentum distribution, and due to its mixing properties, exhibits sample-to-sample autocorrelations that decay far faster than those in the traditional hybrid Monte Carlo algorithm. We test the methods on distributions of varying correlation structure, finding that the proposed technique produces superior covariance estimates, is less reliant on step-size tuning, and can even function with sparse or no momentum re-sampling. The method presented here is promising for more general distributions, such as those that arise in Bayesian learning of artificial neural networks and in the state and parameter estimation of dynamical systems.

\end{abstract}

\maketitle


\section{Introduction}



The Hybrid Monte Carlo (HMC) technique has gained attention over the past few decades as an efficient algorithm for sampling from high-dimensional probability distributions~\cite{duane,neal2010}. An extension of the Metropolis-Hastings Markov Chain Monte Carlo (MCMC) method~\cite{metropolis, hastings}, it has recently shown promise in computational neuroscience and machine learning, where one wishes to sample from high-dimensional and complex posterior distributions that often arise in artificial neural network models~\cite{neal1996, mackay2003,ishwaran}.

HMC builds upon the traditional Metropolis-Hastings method by extending the target distribution to include an auxiliary ``momentum'' distribution, effectively casting the sampling technique as a mechanical system. Samples are collected by motion along the physical trajectories of this system, and due to approximate energy conservation are accepted with high probability, even for well-separated points in the target distribution. HMC has been shown to be quite effective in sampling from a wide array of high-dimensional distributions, often significantly outperforming traditional MCMC~\cite{duane,neal2010}. 

Still, there are issues affecting HMC performance. First, since dynamical motion retains the system on curves of constant energy, and these curves can fail to include criticial sections of the full distribution, momenta variables must often be resampled at every step~\cite{duane, neal2010}. While necessary, this resampling can induce unwanted sample-to-sample correlations, slowing convergence. Another is that the step size of the dynamical integration can reproduce periodicities of the system, effectively returning the same points through many iterations~\cite{neal2010}. Adaptive mechanisms have been developed to confront this issue, but often tests runs may be needed to determine the optimal algorithm specifications~\cite{mackenze}. Various improvements related to these issues and others, such as step-size tuning, alternative integrators, and increasing acceptance rates have been proposed in recent studies~\cite{beskos, CHMC,hmcdetailbalance,shadowhmc,exponentialhmc}.

In this paper, we introduce an adaptation of HMC that exploits the freedom in the choice of momentum distribution to effectively sample from high-dimensional correlated normal distributions along highly mixing, chaotic trajectories. As is characteristic of chaotic systems, autocorrelations decrease exponentially, producing statistics of superior accuracy when compared to traditional HMC for a wide range of algorithm parameters. Importantly, due to the strong mixing properties of the sampling routine, it is more robust to algorithmic fine-tuning than traditional HMC.

The proposed algorithm has been developed at this stage for multivariate normal distributions. Current state-of-the-art sampling techniques of multivariate normals require inversion and QR decomposition of precision matrices, which can be intractable for high-dimensions, particularly when the correlation matrix has little known structure~\cite{highDgaussians}. In that regard, we expect the technique as is to be of immediate relevance. On the other hand, the proposed method sets a template for which to develop similar techniques for more general distributions, particularly those relevant to the estimation of dynamical systems and artificial neural networks.

The paper proceeds as follows. In the following section, we give a review of traditional HMC. Next, we briefly review the notion of the ergodic hierarchy and the motivation for chaotic sampling. The proposed HMC algorithm is then described, followed by results on various correlation matrix structures. Prospects for more general distributions is discussed in the concluding section.


\section{Hybrid Monte Carlo: Overview}
In HMC, the support of the target distribution is extended from the $D$-dimensional space upon which the physical problem is defined, $\bf{X}$, to a $D$+$D$-dimensional space, $\{\bf X, P\}$. The auxiliary ${\bf P}$ variables hold no physical significance, and are only a device to increase minimally-correlated motion in the target space, without sacrificing acceptance probabilty. 

We assume the target distribution can be written as a kernel divided by the normalization factor: $p({\bf X}) = \frac{1}{Z}\Pi({\bf X}) \equiv \frac{1}{Z}\exp{(-U({\bf X}))} $. We assume only that we can evaluate the kernel $\Pi({\bf X})$, or equivalently, the {\it potential energy} $U({\bf X})$ for arbitrary points in the support. 

In HMC, we extend this kernel to $p({\bf X, \bf P}) = \Pi({\bf X})\Xi({\bf P})$, where  $\Xi({\bf P})\equiv \exp(-K({\bf P}))$ is the {\it momentum distribution}. The kernel of the joint density is therefore $\exp({-(K + U))} = \exp{(-E({\bf X,\bf P}))}$, whereby $K$ takes the role of a kinetic energy, $E({\bf X,\bf P})$ assumes the role of the Hamiltonian (often this equals the total energy of the system),  and  ${\bf P}$ are the auxiliary momenta. Since the joint distribution is separable in ${\bf X}$ and ${\bf P}$, then any faithful sampling technique of the joint distribution will independently produce faithful distributions of both, without having to perform any final marginalization. As the ${\bf X}$ variables are the only variables of physical significance, they are retained at the close of the algorithm, while ${\bf P}$ are discarded. 

A tacit but critical requirement of HMC is that sampling from $\Xi({\bf P})$ be simple and inexpensive. The typical form of $\Xi({\bf P})$ is a product of independent Gaussians with zero means and covariances that are either unity or, if possible, some inverse scaling of the corresponding ${\bf X}$ variables~\cite{neal2010}. Being a product of univariate normal distributions, $\Xi({\bf P})$ can be sampled using well-developed techniques~\cite{gaussian}. HMC then uses this as a crutch to sample from the more complex $\Pi({\bf X})$. Indeed, if sampling from $\Xi({\bf P})$ were prohibitive, HMC would add no benefit over traditional Monte Carlo. Though this point is rarely discussed, it will be central in this work.

The joint density endowed with the particular structure just described, the HMC algorithm proceeds as follows. We begin with an initial guess of ${\bf X}_0$, drawn from any distribution, and a guess for ${\bf P}_0$ sampled from its correct distribution, $\Xi({\bf X})$. Next, candidates for the subsequent sample, ${\bf X}_1$ and ${\bf P}_1$ are generated by integrating ${\bf X}_0$  and ${\bf P}_0$ forward using Hamilton's equations of motion derived from the function $E({\bf X}, {\bf P})$~\cite{goldstein}:
\begin{align}
\dot {\bf X} &= \frac{dE}{d{\bf P}} \nonumber \\
\dot {\bf P} &= -\frac{dE}{d{\bf X}} 
\label{eq:ham}
\end{align}
Integrating this set of differential equations from (${\bf X}_0, {\bf P}_0)$ for some finite time gives the candidate point (${\bf X}^*, {\bf P}^*$). The candidate is accepted or rejected based on the ratios of probabilities between steps, $\pi = {\exp(-(E({\bf X}^*, {\bf P}^*)-E({\bf X}_0, {\bf P}_0)))}$, much like the standard Metropolis algorithm, but using the joint distribution instead of the target distribution alone. The candidate is accepted with probability $\min (1,  \pi)$. Next, the momenta is again resampled from its correct distribution, and the algorithm repeats. 

The $D$+$D$-dimensional support $\{ {\bf X}, {\bf P}\}$ is a {\it symplectic manifold}, preserving key quantities along any integral curve of Hamilton's equations, Eqs. \ref{eq:ham}~\cite{goldstein, hairer}. One of these is the Hamiltonian $E$, which results in high acceptance probabilities. Another is phase space area, which means that the distribution is not deformed by HMC algorithm. Clearly the first criterion is desirable, while the second is necessary~\cite{neal2010} (this latter condition can actually be relaxed by introducing appropriate scale factors, see~\cite{shadowhmc,reichnosehoover}).

For practical cases of interest, the differential equations of motion cannot be solved analytically, and propagating ${\bf X}_0$ and ${\bf P}_0$ forward requires numerical integration. This introduces errors into the estimates of ${\bf X}$ and ${\bf P}$, and therefore deviations of the system from its true trajectory. 
These errors can accumulate without bound, 
even for numerical integration of analytically solvable systems such as the harmonic oscillator~\cite{hairer}. For physical systems, this implies that the estimate of the system degrades appreciably over long integration times. For HMC, it implies that phase space areas will shrink or dilate as more samples are collected, deforming the target distribution. In other words,  symplectic symmetry is lost when discretizing Hamilton's equations.

There are, however, a class of integrators known as symplectic, or geometric integrators, which bound the errors on ${\bf X}$ and ${\bf P}$ at every step, such that these errors do not accumulate in time~\cite{hairer}. Furthermore, they preserve phase space areas at each step exactly. 
Many common integrators are not symplectic; among these are the first-order Euler scheme, the second-order trapezoidal rule, and the fourth-order Runge-Kutta method. Others, such as the midpoint rule, are symplectic. Symplectic integrators to arbitrary order, of differing schemes, have been systematically constructed over the years~\cite{hairer}. The second-order symplectic leapfrog, or Verlet, integrator is used almost universally in HMC algorithms and variations thereof~\cite{neal2010}. 

The benefit of the HMC sampling technique is its large reduction in sample correlation. In traditional Monte Carlo, the sampler moves from point to point using random walks, so searches can take considerable time to adequately explore the state space. In constrast, HMC allows the exploration of distant points ${\bf X}$ by also allowing motion in the fictitious ${\bf P}$ space; however, it retains high acceptance probability by moving along curves of nearly constant energy, which is guaranteed for Hamiltonian flows. The caveat ``nearly''  arises from the fact that the discretization errors spoil the exact preservation of the energy, whereby the rejection criteria is needed. Nonetheless, as the Verlet scheme is a second-order integrator, the acceptance criteria can still be near unity up to appreciable step sizes.


\section{Ergodicity, Mixing and Correlations}

Proposal samples in HMC are gathered by numerically integrating Hamilton's equations forward in time. Though this motion can more effectively search state space than random walks alone, there are gradations to how ``effective'' this search can be. can As a simple illustration, consider first applying the HMC algorithm to a 2$D$ Gaussian distribution with diagonal covariance matrix: $\Sigma_{11} = \sigma_1^2; \Sigma_{22} = \sigma_2^2$. For this distribution the Hamiltonian is:
\begin{align*}
E({\bf X, \bf P}) &=\frac{x_1^2}{2\sigma_1^2} + \frac{x_2^2}{2\sigma_2^2} + \frac{p_1^2}{2} + \frac{p_2^2}{2} 
\end{align*}
Using Hamilton's equations for this system, the motion is that of two uncoupled oscillators of unequal frequency:
\begin{align*}
\dot x &=p_x \\ 
\dot y &= p_y \\
\dot p_x &=  -x/\sigma_1^2 \\
\dot p_y &= -y/\sigma_2^2 \\ \\
\rightarrow x(t) &\sim \cos(t/\sigma_1)  \\
\rightarrow y(t) &\sim \cos({t/\sigma_2}) 
\end{align*}
The motion of this oscillator in the $xy$ or $xp_x$ plane depends upon the relationship between the frequencies $1/\sigma_1$ and $1/\sigma_2$. If the ratio is rational, the curve generated will execute a repeating motion that closes upon itself indefinitely~\cite{goldstein}. For irrational ratios, the frequencies are incommensurate, and the motion will arrive arbitrarily close to any point on the energy manifold; this is known as ergodic motion, or quasiperiodicity~\cite{goldstein}. 



The idea that the dynamical trajectory can traverse arbitrarily close to another point in the support suggests that ergodic systems densely sample the target distribution. But from the standpoint of randomness, ergodicity alone is a relatively weak condition. Other types of dynamical motion can produce behavior that is in a sense ``more random'' than ergodicity. 

A precise quantification of these ideas is encapsulated in a classification known as the ergodic hierarchy. Though there is some disagreement about whether the rungs of this hierarchy can really be defined in terms of ``randomness'' {\it per se}~\cite{ergodichierarchy}, there is nevertheless a well-defined delineation based on the behavior of the correlations under forward mappings~\cite{ergodichierarchy,mixingbook}. For our purposes, the correlation is defined as the sample autocorrelation:
\begin{align}
C(\tau) &= \lim_{N\rightarrow \infty}\sum_k^N f(t_k - \tau)f(t_k) 
\end{align}

\begin{figure}[tb]
\centering
\includegraphics[height=2.85cm]{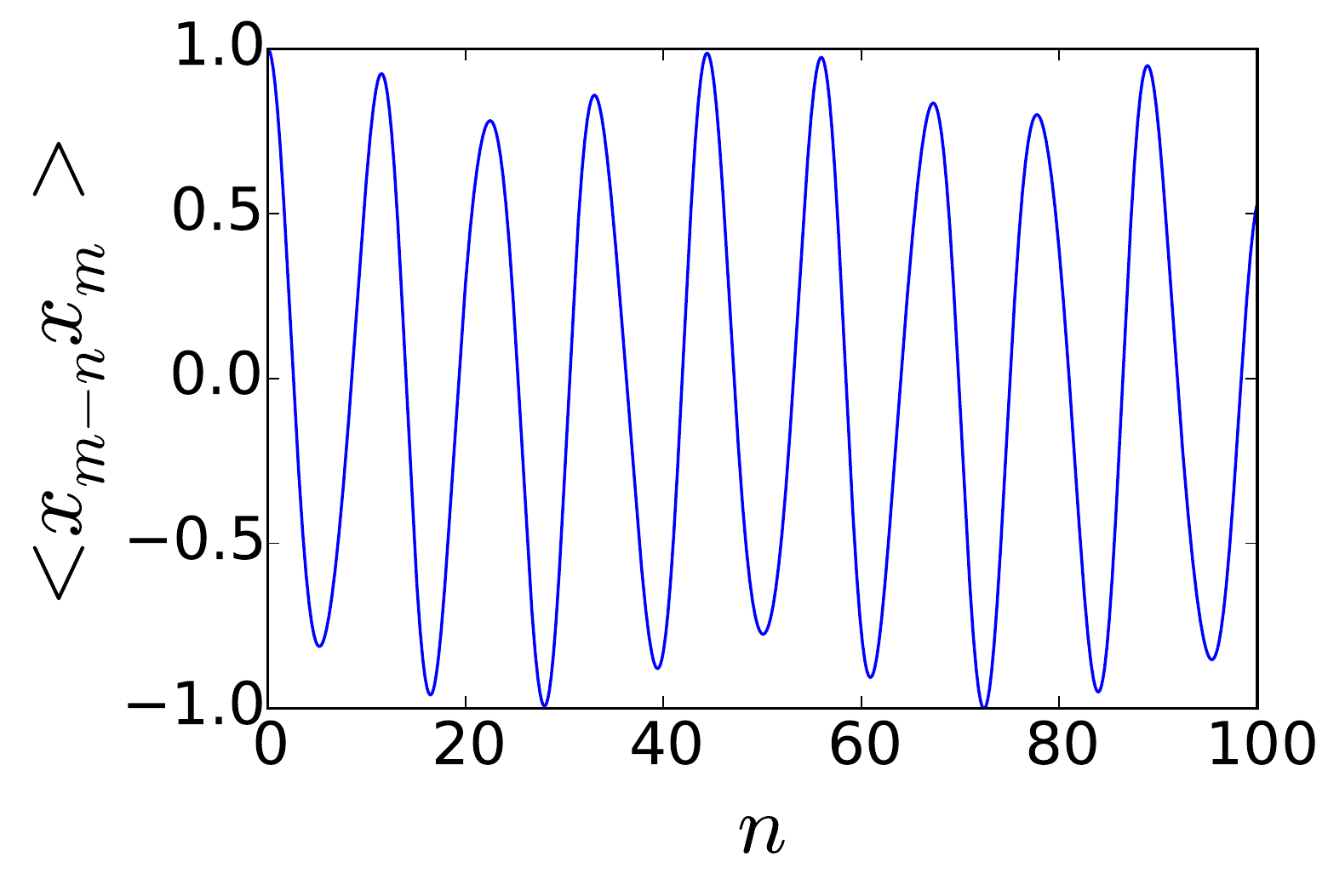}
\includegraphics[height=2.85cm]{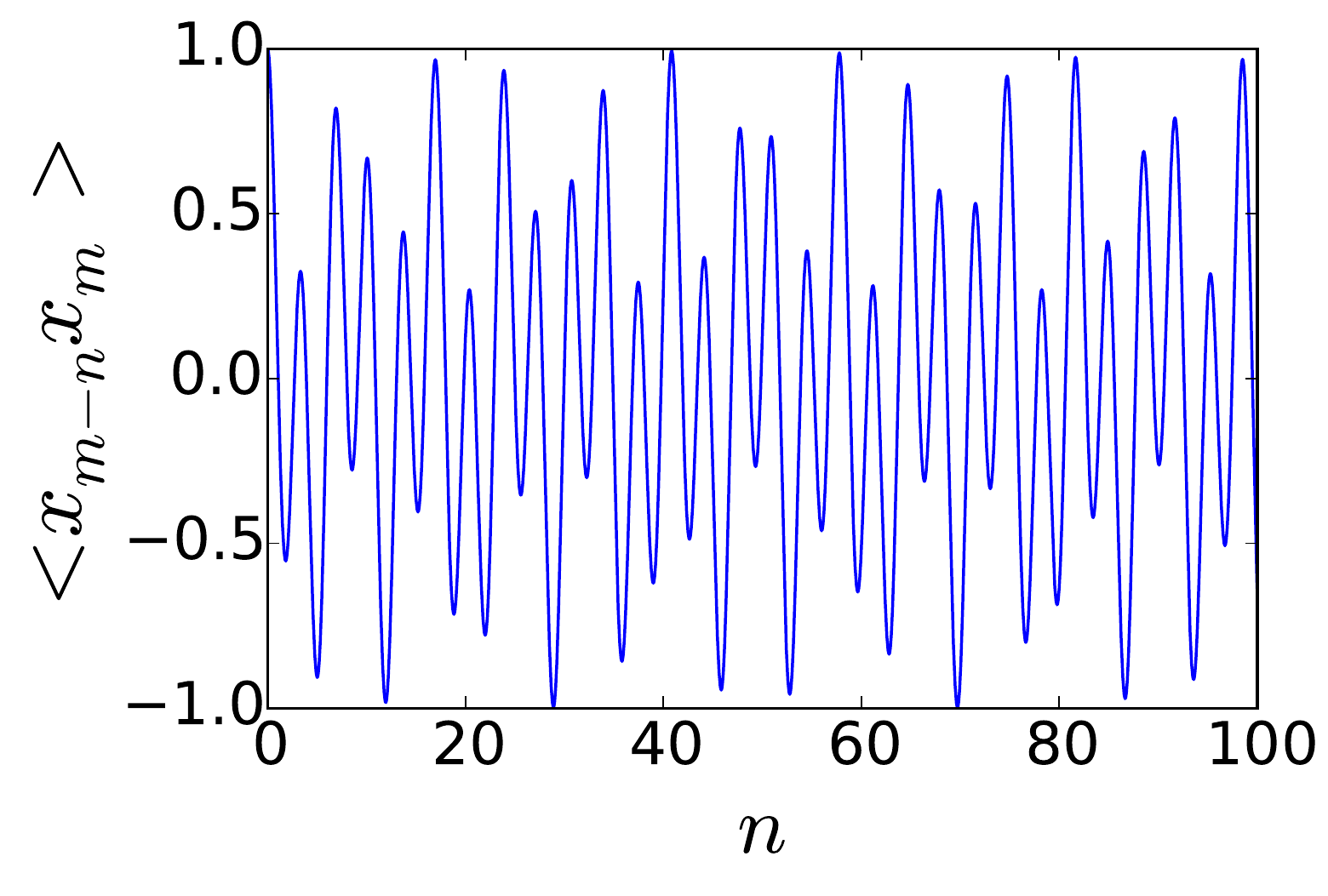}
\caption{{\it Left} Autocorrelation function for 2$D$ uncoupled quasiperiodic oscillator; $\omega_2/\omega_1 = \pi$. {\it Right} Autocorrelation function for a coupled 2$D$ oscillator.}
\label{fig:2DHOcorr}
\end{figure}

Ergodic systems occupy the lowest rung of the ergodic hierarchy. They are characterized by the equivalence of time averages and spatial averages~\cite{ergodichierarchy, mixingbook}. The intuition behind this is that, for a long enough time, a single dynamical trajectory produces the same statistics as an ensemble of samples. However, ergodic systems may exhibit long-lived correlations which can make the time for these statistics to converge unacceptably long; they do not necessarily require $C(\tau)$ itself approach zero asymptotically. As an example, we show the autocorrelation function $C(\tau)$ for the uncoupled 2$D$ oscillator with $\omega_1/\omega_2 = \pi$, in Figure \ref{fig:2DHOcorr}. For comparison, we also show in Figure \ref{fig:2DHOcorr} the autocorrelation function for a 2$D$ {\it coupled} oscillator with 

\begin{align*}
E({\bf x, \bf p}) &=\frac{x_1^2}{2}  + \frac{x_2^2}{2} - \frac{x_1x_2}{\sigma_{12}}+ \frac{p_1^2}{2} + \frac{p_2^2}{2} \end{align*}

where $\sigma_{12} = 1/\sqrt{2}$. Though these systems are both ergodic, significant autocorrelations persist indefinitely.

Above ergodicity on the ergodic hierarchy are ``mixing'' systems~\cite{ergodichierarchy, mixingbook}. In mixing systems, correlations necessarily decay to zero:
\begin{align}
\lim_{\tau \rightarrow \infty} \sum_\tau C(\tau) &= 0
\label{eq:mixingcriterion}
\end{align}
Many mixing systems can also be classified as chaotic, characterized by the existence of positive Lyapunov exponents, which quantify the speed at which nearby trajectories diverge in time~\cite{goldstein, ott}. For mixing and chaotic systems, it is generally the case that the autocorrelations not only die to zero, but do so at an exponential rate (this distinction is not entirely clear-cut; as such, the correlation decay rate is often studied for particular classes of systems on an individual basis~\cite{ergodichierarchy, baladi2000}). 
If the Hamiltonian dynamics in an HMC sampler were truly mixing, we would expect a strong decay of autocorrelations arising from the dynamical step alone. Such a sampler would be far less sensitive to the correlating effects of momentum re-sampling, as well as to ill-chosen initializing guesses of the routine; in traditional routines this latter effect necessitates a ``burn-in'' period~\cite{neal2010}.


\section{Inducing Chaos Via Non-Canonical Kinetic Energy Terms}
 
 \subsection{(2+2) Dimensions}
 
\begin{figure}
\includegraphics[height=4.2cm]{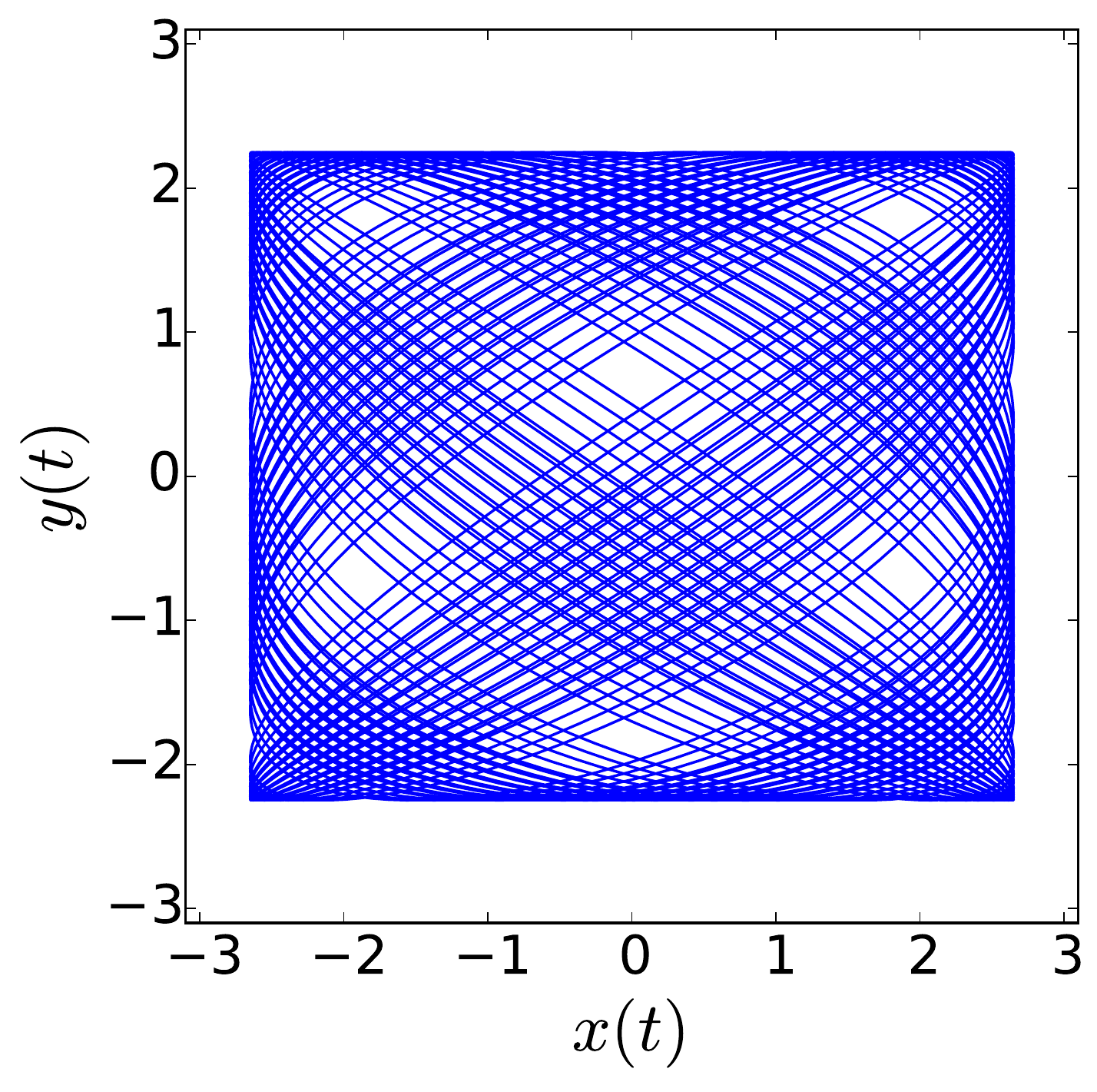}
\includegraphics[height=4.2cm]{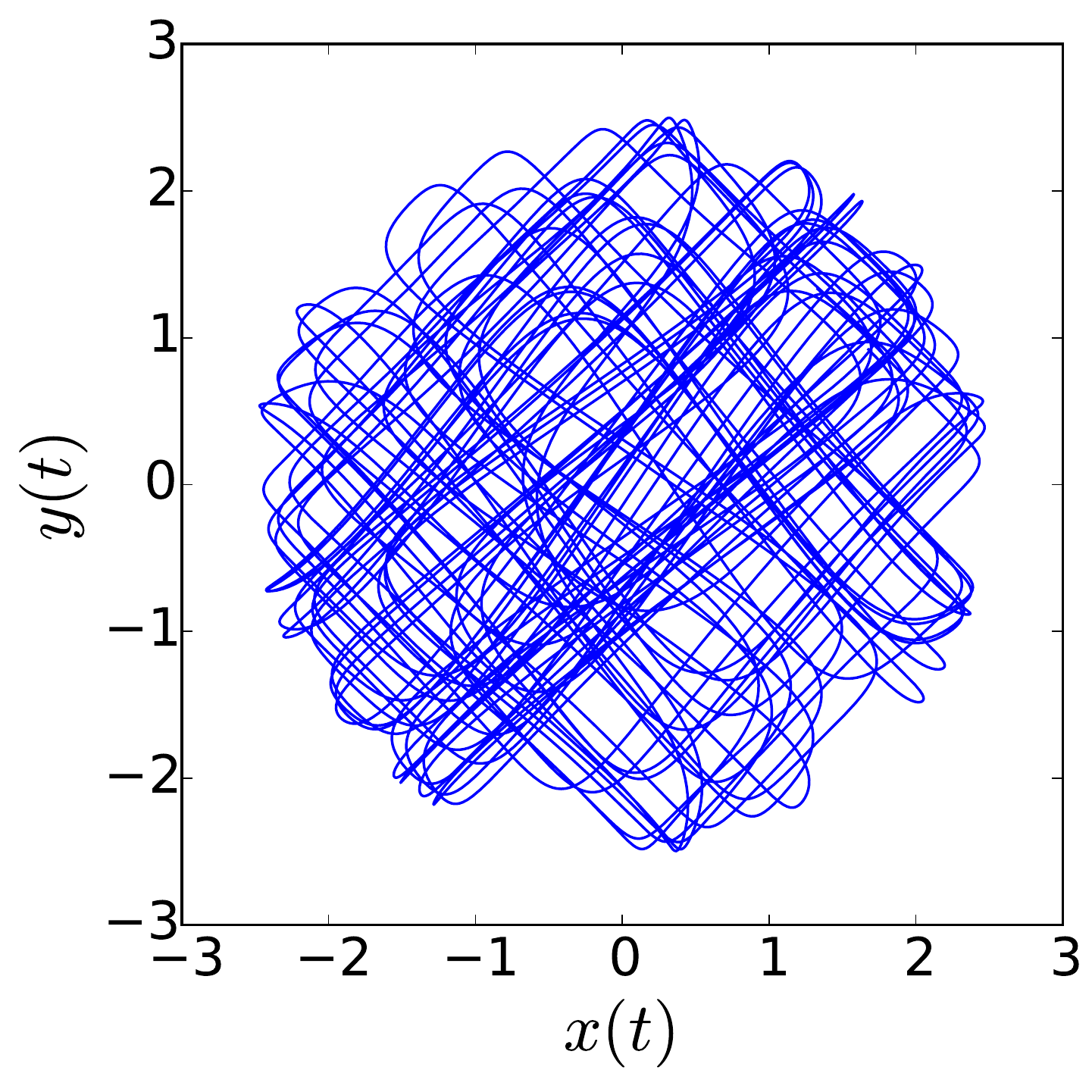}
\includegraphics[height=4.7cm]{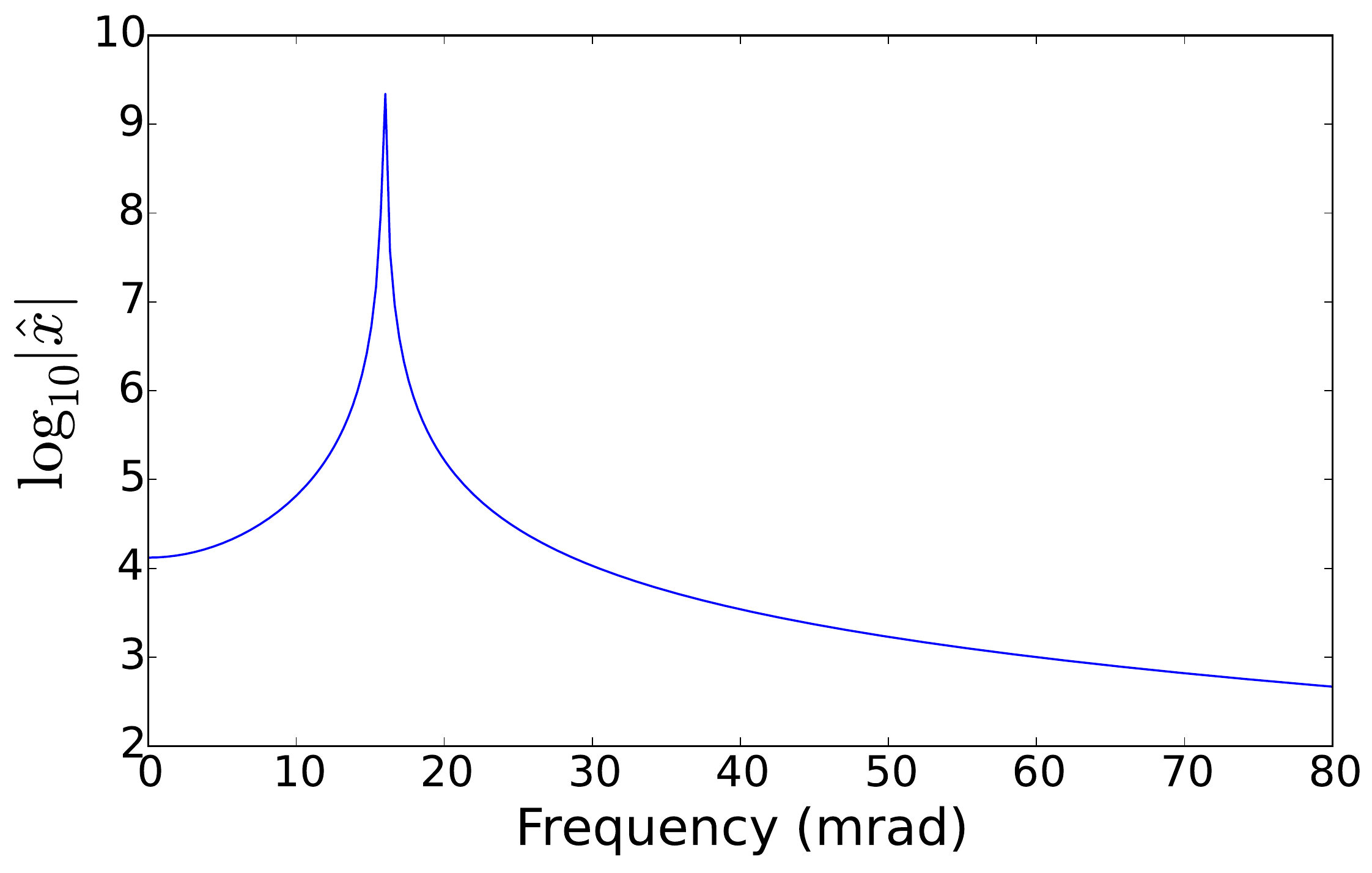} \\
\includegraphics[height=4.7cm]{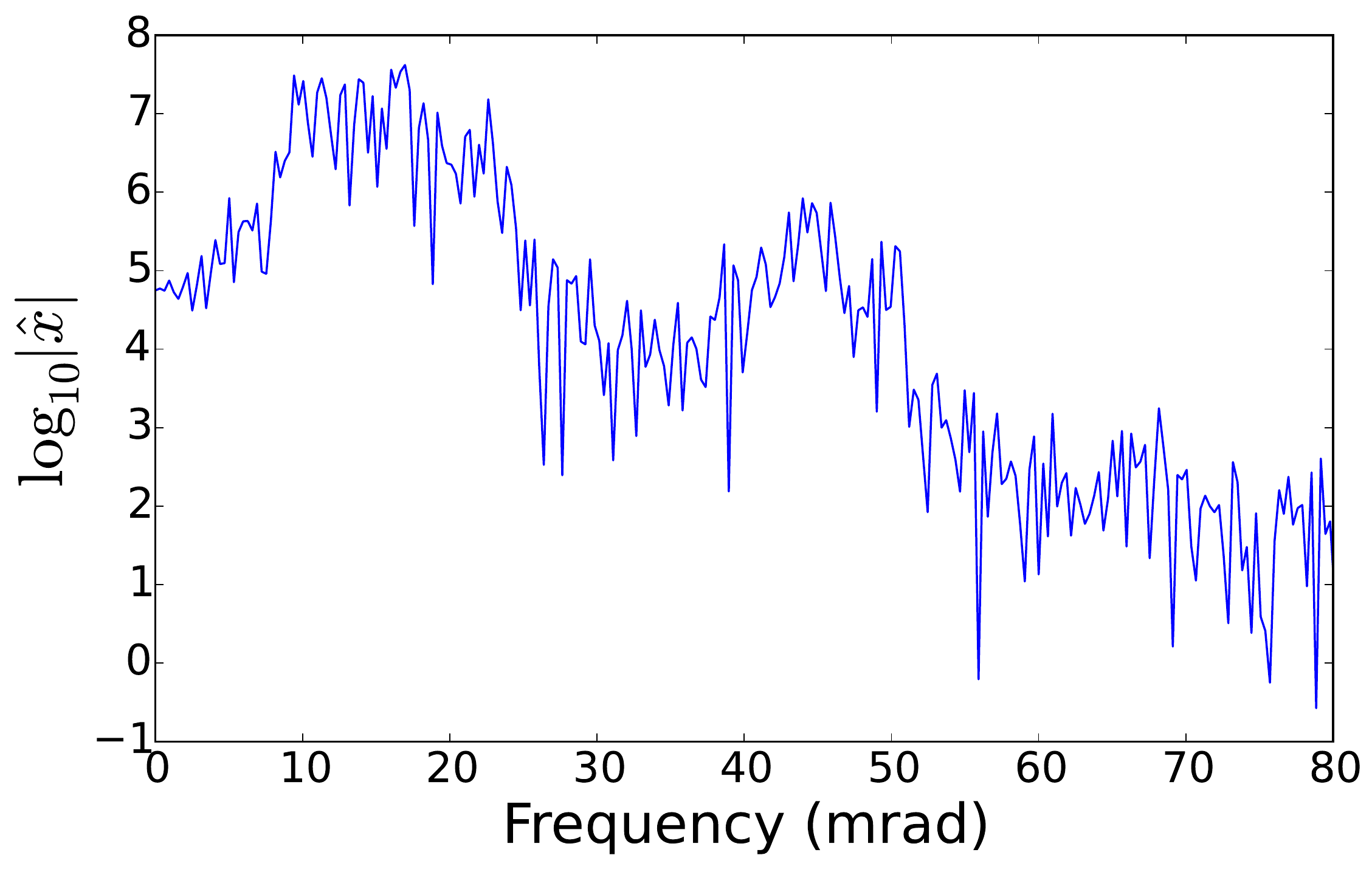}
\caption{{\it Top Row} Typical phase plots in the $x$-$y$ plane of the (2+2)$D$ uncoupled oscillator (left) and the (2+2)$D$ Hamiltonian chaotic system (right). The lower plots show the logarithm of the Fourier transforms of $x$ for these two systems, respectively. The broad continuous component in the lowest plot is a signature of chaotic dynamics.}
\label{fig:Phase_Plot}
\end{figure}

To this end, we point to the freedom in the choice of $\Xi({\bf P}) = \exp(-K({\bf P}))$. Indeed, since the potential energy is fixed from the target distribution, any attempt at producing different dynamics will necessarily have to be done either with added position variables or via the momentum alone. 

To fix ideas, we first consider sampling from a 2$D$ Gaussian with covariance $ \mathbf \Sigma = {\mathbf I}$. There are many choices of $K({\bf P})$ which exhibit chaotic regimes~\cite{elegantchaos}, but there is benefit in simplicity. The following choice is ideal:
\begin{align*}
U({\bf X}) &=  x^2/2\sigma_x^2 + y^2/2\sigma_y^2 \\
K({\bf P}) &= p_x^2\sigma_x^2/2  + p_y^2\sigma_y^2/2 + p_x^2p_y^2\sigma_x^2\sigma_y^2/2 
\end{align*}
The system is chaotic for a large section of initial conditions~\cite{elegantchaos}. A typical trajectory in $x$-$y$ space is plotted in Figure \ref{fig:Phase_Plot}, alongside the discrete Fourier transform  (DFT) of the first variable. As is indicative of chaotic systems, the absolute value of the DFT exhibits a broad continuous component. For this particular example, the global Lyapunov exponents, which we calculated using the Oseledec Theorem~\cite{nonlineartimeseries}, are $\{-0.013,. 0.018, 0.439, -0.437\}$. As for all Hamiltonian systems, they exist in plus-minus pairs; this is a consequence of the invariance of phase space area. Importantly, the maximal exponent is non-negligible and positive, indicating chaotic dynamics \cite{ott}. As a comparison, the phase plot and DFT for a canonical momentum distribution, $p_x^2/2  + p_y^2/2$,  are also shown in Figure \ref{fig:Phase_Plot}. 

In HMC, the computational benefit of the quadratic kinetic energy is that the momentum randomization step amounts to sampling from a univariate normal distribution, for which dedicated quick and efficient random number schemes exist~\cite{gaussian}. On the other hand, no such guarantee exists for an arbitrary kinetic energy of high dimension; if the sampling routine is more expensive than the traditional Monte Carlo method, or cannot effectively scale to higher dimensions, then the benefit of HMC has been lost.

Fortunately, for the system described, momenta can be sampled with the acceptance-rejection method~\cite{montecarlomethods}. The 2$D$ kinetic energy of this chaotic system is bounded below by the separable quadratic function $\sigma_x^2p_x^2/2 + \sigma_y^2p_y^2/2$, whereby the kernel $\Xi({\bf P})$ is bounded above by product of two univariate Gaussian kernels of variance $1/\sigma_1^2$ and $1/\sigma_2^2$, respectively. This product therefore give a legitimate proposal distribution~\cite{montecarlomethods}. We calculate the rejection rate by dividing the integrals of the target kernel $\Xi({\bf P})$ and the proposal kernel, $\Xi^*({\bf P}) \equiv \exp(-\sigma_x^2 p_x^2/2-\sigma_y^2 p_y^2/2)$:
\begin{align*}
\int \Xi({\bf P}) d {\bf P}&\approx \frac{2.70141}{ \sqrt{\sigma_x\sigma_y}}  \\
\int \Xi^*({\bf P}) d {\bf P}&=  \frac{\pi}{ \sqrt{\sigma_x\sigma_y}}
\end{align*}

The acceptance rate is $2.7/\pi =  85.9\%$, so that only two proposals are necessary to adequately sample from $\Xi({\bf P})$.

 One common limitation of rejection sampling is that acceptance rates fall exponentially with dimension~\cite{montecarlomethods}. Fortunately, we can easily adapt this particular coupling for arbitrarily high dimensions as we now discuss.


\subsection{Arbitrary Dimensions}
We now consider the general problem. We wish to sample from a Gaussian distribution of dimension $D$ and known precision matrix ${\mathbf \Sigma^{-1}}$. Assume for the time being that $D$ is even. Guided by 2$D$ example, we propose the following kinetic energy function:
\begin{align*}
K({\bf P}) &= \sum_{i \textup { even}}^{D} \frac{1}{2\Sigma^{-1}_{i,i}}p_{i}^2 + \frac{1}{2\Sigma^{-1}_{i+1,i+1}}p_{i+1}^2\\
&+ \frac{1}{2\Sigma^{-1}_{i,i}\Sigma^{-1}_{i+1,i+1}}p_{i}^2 p_{i+1}^2 
\end{align*}

This form effectively generalizes the 2$D$ momenta coupling to $D/2$ pairs. This particular form also inversely scales ${\mathbf P}$ to ${\bf X}$ to an appreciable degree, without having to explicitly invert the precision matrix. This prevents massive differences of scale between the $\dot {\mathbf P}$ and the $\dot {\mathbf X}$  equations, which can push the system out of the chaotic regime. Finally, the momentum resampling can be done as proposed above, via rejection sampling upon univariate Gaussian proposal distributions:
\begin{align*}
\Xi^*({\bf P}) = \prod_i^{D}\exp(-p_i^2/2\Sigma^{-1}_{i,i}) 
\end{align*}

Since each pair of momenta is sampled separately with acceptance probability 86\% per pair, the number of ${\mathbf P}$ samples needed to randomize the momentum scales only linearly with the dimension, i.e. $\sim D$ samples from a univariate Gaussian. 

If the target distribution is defined in an odd number of dimensions, the pairwise momenta coupling presented above can be used for any $D-1$ of the dimensions, while the kinetic energy term of the remaining term is chosen as the usual uncoupled kinetic energy term:
\begin{align*}
K(p_{D}) &= \frac{1}{2\Sigma^{-1}_{D,D}}p_D^2
\end{align*}

Though this last term is on its own non-chaotic, we expect that for high-dimensional distributions the coupling of the last dimension to the others via correlations in the target multivariate normal will nonetheless induce high mixing in the last dimension as well.


\section{Methods}


\subsection{Covariance and Precision Matrices}

We tested these ideas on 100-dimensional multivariate normal distributions of varying correlation structure: either uniformly random correlations, or Toeplitz-type correlations. 

For  uniform correlations, covariance matrices were generated by choosing the elements of a $D$x$D$ matrix from a uniform distribution between some $|\alpha| < 1$. For 100 dimensions, $\alpha$ was about 0.15. This matrix, $\mathbf \Sigma$, is then symmetrized via $\mathbf \Sigma \rightarrow \frac{1}{2}(\mathbf \Sigma + \mathbf \Sigma^T)$ and its diagonals were set to unity. Its eigenvalues were calculated to ensure positive semi-definiteness; if not, the process was repeated. Finally, it was inverted to create the precision matrix, $\mathbf \Sigma^{-1}$. 

For Toeplitz-type correlations, we generated both ``linear'' and ``geometric'' types. Geometric Toeplitz matrices were built beginning with:
\begin{align*}
\mathbf A = 
\left| 
\begin{array}{cccccc}
1 & \alpha & \alpha^2 & ... & \alpha^{D-2} & \alpha^{D-1} \\
\alpha & 1 & \alpha & \alpha^2 &...&\alpha^{D-2} \\  
. &.&&&&.\\
. &&.&&&.\\
. &&&.&&.\\
\alpha^{D-2}&&&\alpha&1 & \alpha\\
\alpha^{D-1} & \alpha^{D-2} & ...& \alpha^2 & \alpha & 1 \\  
\end{array}
\right|
\end{align*}

where $\alpha$ is chosen uniformly between -1 and 1. $\mathbf A$ was slightly randomized via:
\begin{align*}
\mathbf A \rightarrow \mathbf A\cdot \mathcal{N}(1,\alpha/3)
\end{align*}

Next, ${\mathbf A}$ was symmetrized and its diagonal was reset to unity to generate $\mathbf \Sigma$. Its eigenvalues were determined to ensure that it retained positive semi-definiteness following the multiplicative randomization. Finally $\mathbf \Sigma$ was inverted to generated the precision matrix $\mathbf \Sigma^{-1}$. 

Linear Toeplitz matrices were generated similarly, but instead beginning with:

\begin{align*}
\mathbf A = 
\left| 
\begin{array}{cccccc}
1 & \alpha & \frac{\alpha}{2} & ... & \frac{\alpha}{D-2} & \frac{\alpha}{D-1} \\
\alpha & 1 & \alpha & \frac{\alpha}{2} &...&\frac{\alpha}{D-2} \\  
. &.&&&&.\\
. &&.&&&.\\
. &&&.&&.\\
\frac{\alpha}{D-2}&&&\alpha&1 & \alpha\\
\frac{\alpha}{D-1} & \frac{\alpha}{D-2} & ...& \frac{\alpha}{2} & \alpha & 1 \\  
\end{array} 
\right| \\
\end{align*}

Given any of these three correlation structures, the target kernel to be sampled is:
\begin{align*}
\Pi({\bf X}) &= \exp(-E({\bf X})) \\
&= \exp(-{\bf X} \mathbf \Sigma^{-1} {\bf X}/2) = \exp(-x_i  \Sigma_{ij}^{-1} x_j/2) 
\end{align*}
 

\subsection{Traditional HMC with Scaling}
Traditional hybrid Monte Carlo samples were generated using the following momentum kernel, which is a product of $D$ univariate Gaussians:

\begin{align*}
\Xi({\mathbf P}) =\prod_i^{D}\exp\left(\frac{1}{2\Sigma^{-1}_{i,i}}p_{i}^2\right)
\end{align*}

The scaling factors $\Sigma^{-1}_{i,i}$ are known to improve the performance of traditional HMC particularly for ill-conditioned covariance matrices~\cite{neal2010}. Here too we have found that the scaled form vastly outperforms the unscaled kernel $K({\bf P}) = \sum_i^{D}\exp(-p_i^2/2)$, so we will use the scaled form only for comparison against the proposed method. 

Traditional HMC proceeds as follows. It initializes as the following:
\begin{enumerate}
\item Choose ${\mathbf X}_0$ from some prior distribution.
\item Sample ${\mathbf P}_0$ from $\Xi({\mathbf P})$, equivalent to sampling from $D$ univariate Gaussians with zero mean and variances $\Sigma^{-1}_{1,1},\Sigma^{-1}_{2,2},...,\Sigma^{-1}_{D,D}$.
\end{enumerate}

The following steps are then repeated iteratively for $N-1$ steps:

\begin{enumerate}
\item Generate new proposal samples ${\bf X}^*$ and ${\bf P}^*$ by numerically integrating Hamilton's equations forward from $\{{\bf X}_n$,${\bf P}_n\}$ for $L$ steps, using a Stormer-Verlet symplectic integrator with a timestep of $h$.
\item Calculate $\exp(\Delta E) \equiv \exp(-K({\bf P}^*) - U({\bf X}^*)+ K({\bf P}_n) +  U({\bf X}_n))$. Sample $\gamma$ uniformly from $[0,1]$. If $\gamma < \exp(-\Delta E) $, then set $\mathbf X^* \rightarrow {\bf X}_{n+1}$; otherwise, set $\mathbf X_n \rightarrow {\bf X}_{n+1}$. This is equivalent to accepting $\{{\bf P}^*, {\bf X}^* \}$ with probability $\min (1,\exp(-\Delta E))$.
\item Randomize the momenta by sampling ${\mathbf P}_{n+1}$ from $\Xi({\mathbf P})$.
\end{enumerate}

$K$ such routines are run in parallel, and the pooled $N\cdot K$ samples from these runs are then used to calculate statistics, which is compared to the generating correlation matrix $\mathbf \Sigma$ to assess the accuracy of the results.


\subsection{HMC with Chaotic Mixing}
The HMC algorithm proposed in this paper proceeds in a similar fashion. The difference lies in the momentum kernel, which is:

\begin{align*}
\Xi({\mathbf P}) = \prod_{i \textup { even}}^{D} &\exp\left(\frac{1}{2\Sigma^{-1}_{i,i}}p_{i}^2\right)\exp\left(\frac{1}{2\Sigma^{-1}_{i+1,i+1}}p_{i+1}^2 \right)\\
\times&\exp\left(\frac{1}{2\Sigma^{-1}_{i,i}\Sigma^{-1}_{i+1,i+1}}p_{i}^2 p_{i+1}^2 \right) \\
\equiv \prod_{i \textup{ even}}&\Xi_i(p_i,p_{i+1}) 
\end{align*}

The algorithm is initialized as follows:
\begin{enumerate}
\item Choose ${\mathbf X}_0$ from some given (prior) distribution.
\item $[$Rejection sampling$]$ For each even dimension $i$, individually sample $p_i$ and $p_{i+1}$ from univariate Gaussian distributions of mean zero and variance $\Sigma^{-1}_{i,i}$ and $\Sigma^{-1}_{i+1,i+1}$, respectively. Then, use the chosen $p_i$ and $p_{i+1}$ to sample $\gamma_i$ uniformly from $[0,\exp(-p_i^2/2\Sigma_{i,i}- p_{i+1}^2/2\Sigma_{i+1,i+1})]$. Accept $p_i, p_{i+1}$ if $\gamma < \Xi_i(p_i,p_{i+1})$, i.e. if $\gamma$ lies within the envelope of the pdf of the target distribution at the sampled value $p_i,p_{i+1}$. If rejected, repeat again. Since the acceptance ratio probability is $85.6\%$, only 2-3 samples per even dimension $i$ will be needed to produce an accepted set of $p_i,p_{i+1}$ (this can be done in parallel). This sampling step produces ${\mathbf P}_0$.
\end{enumerate}

The same steps as in the traditional HMC are then iterated $N-1$ times, using instead the form of the momentum kernel indicated here. The momentum randomization at the end of each step is performed using the same rejection sampling routine as noted in the initialization step. We will refer to this method as ``Chaotic Hybrid Monte Carlo,'' or ``CHMC.''


\subsection{Calculation Details}
The efficacy of CHMC was tested by comparing the statistics of the samples against a typical, scaled hybrid Monte Carlo algorithm. We define a single run as a HMC or CHMC sampling by $K=100$ initial conditions (``walkers'') upon a 100-dimensional Gaussian target kernel, defined as:
\begin{align*}
\Pi({\mathbf X}) = \exp(-\mathbf X \mathbf \Sigma^{-1} \mathbf X) 
\end{align*}

We use $N=2000$, $L=50$, and $h \in \{0.01, 0.05, 0.1, 0.15, 0.2$, or $0.25\}$. The estimated covariance matrix $\hat {\mathbf \Sigma}(n)$ was calculated as a function of sample step $n$ by calculating the covariance of all $n\cdot K$ samples up to that sample. This allows us to see how the rate at which estimates improve as more samples are taken. We then compared this to the actual covariance of the target kernel by considering the mean squared errors of the off-diagonal and on-diagonal elements separately:
\begin{align*}
\textup{MSE\textsubscript{off}}(\hat {\mathbf \Sigma}(n)) &= \frac{ \sum_{i \ne j} ( {\hat{\mathbf \Sigma}_{ij}}(n) - {\mathbf \Sigma}_{ij})^2}{D(D-1)}\\
\textup{MSE\textsubscript{on}}(\hat {\mathbf \Sigma}(n)) &= \frac{\sum_{i} ( {\hat{\mathbf \Sigma}_{ii}}(n) - {\mathbf \Sigma}_{ii})^2}{D}
\end{align*}

Finally, this was averaged over 50 different matrices for each of the three types of correlation matrices considered.


\section{Results and Discussion}

\begin{figure*}
\centering
\includegraphics[width=5.8cm]{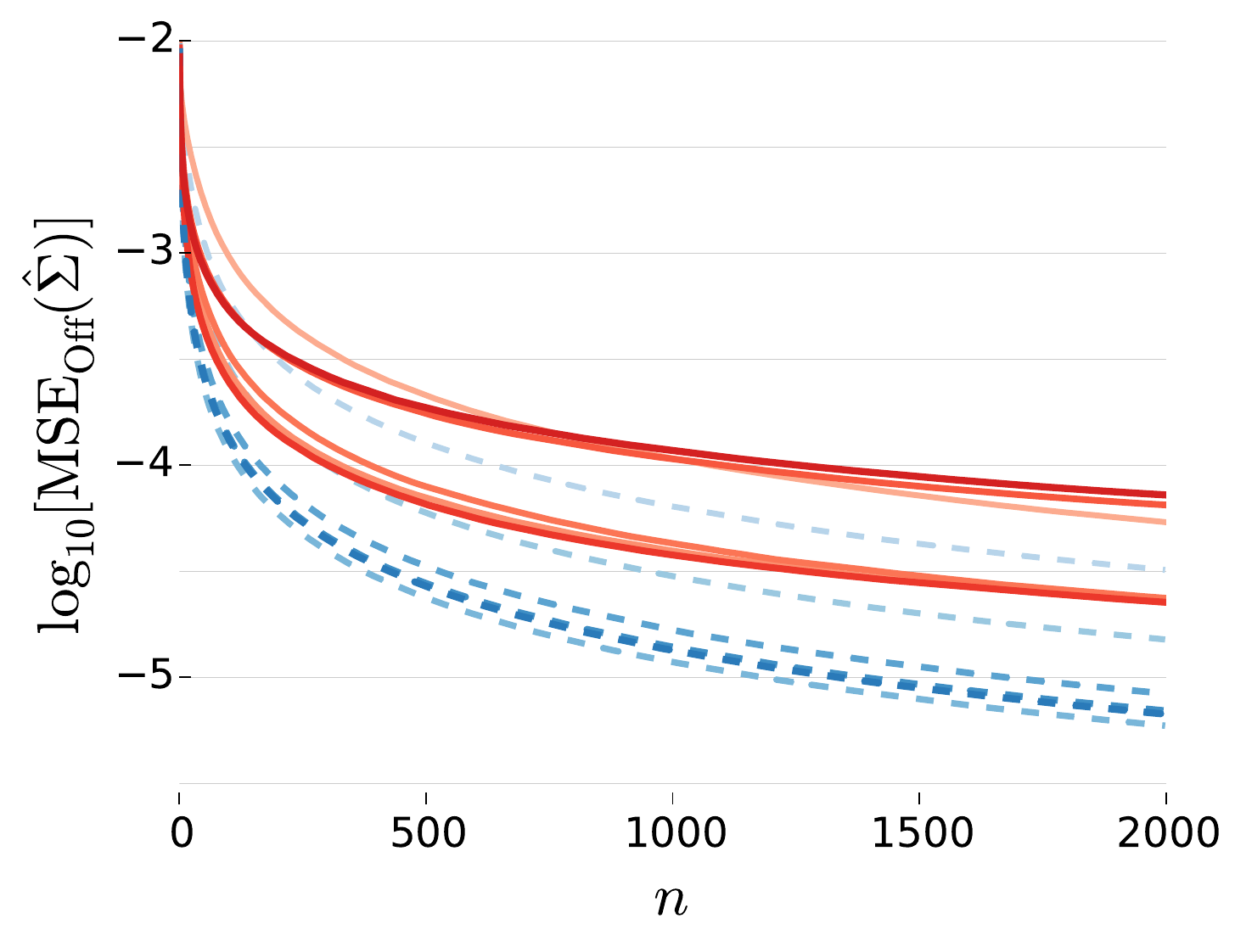}
\includegraphics[width=5.8cm]{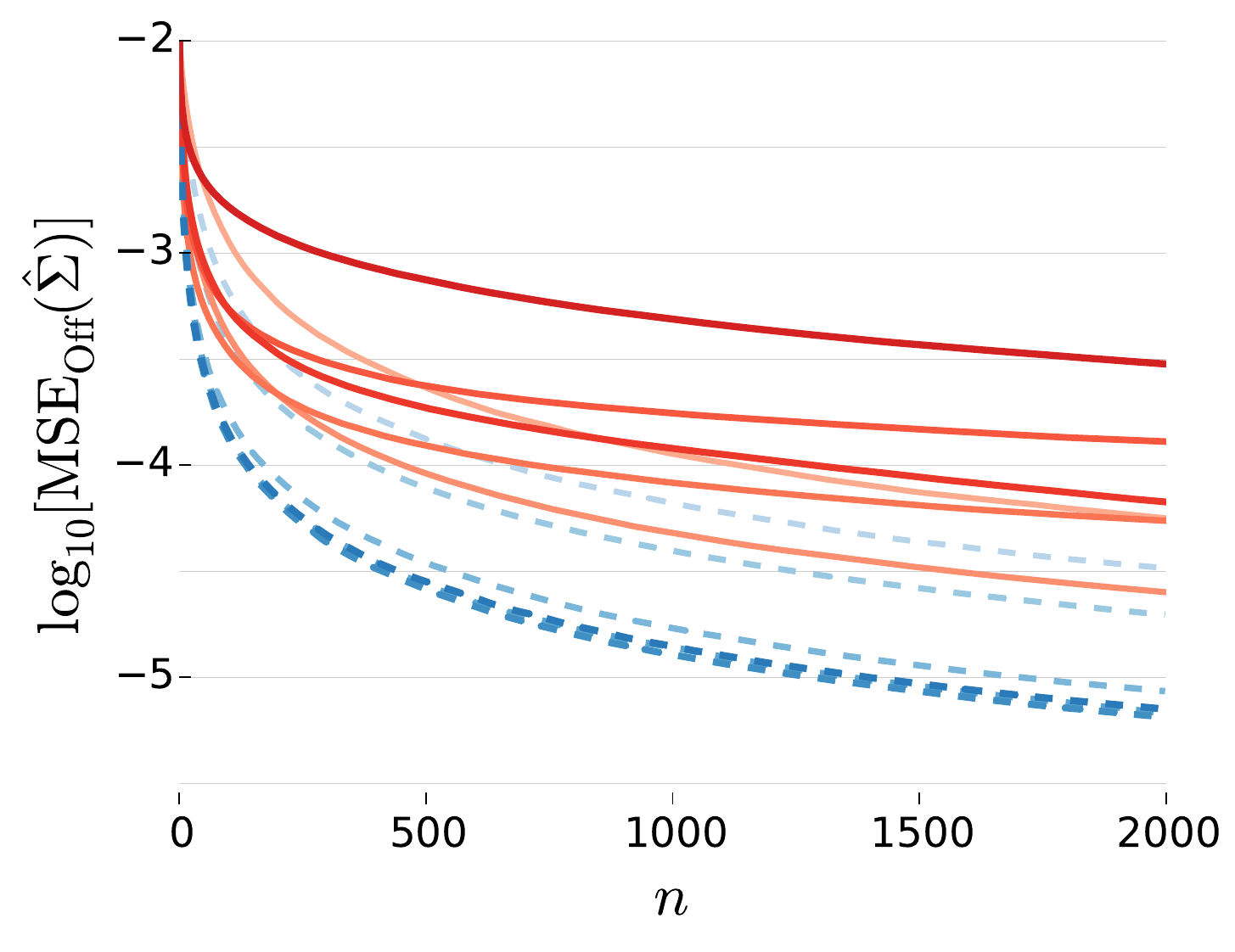}
\includegraphics[width=5.8cm]{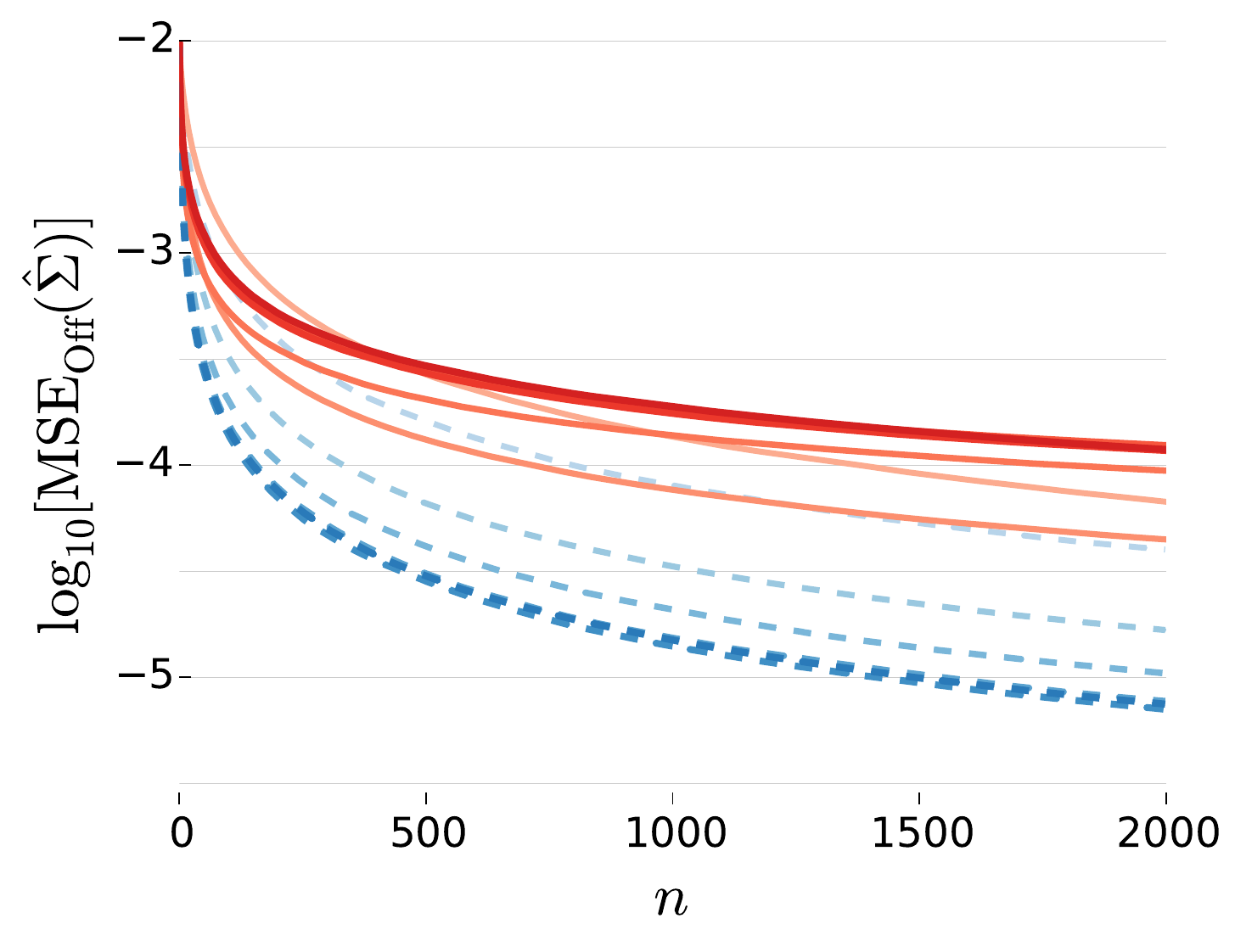} \\
\includegraphics[width=5.8cm]{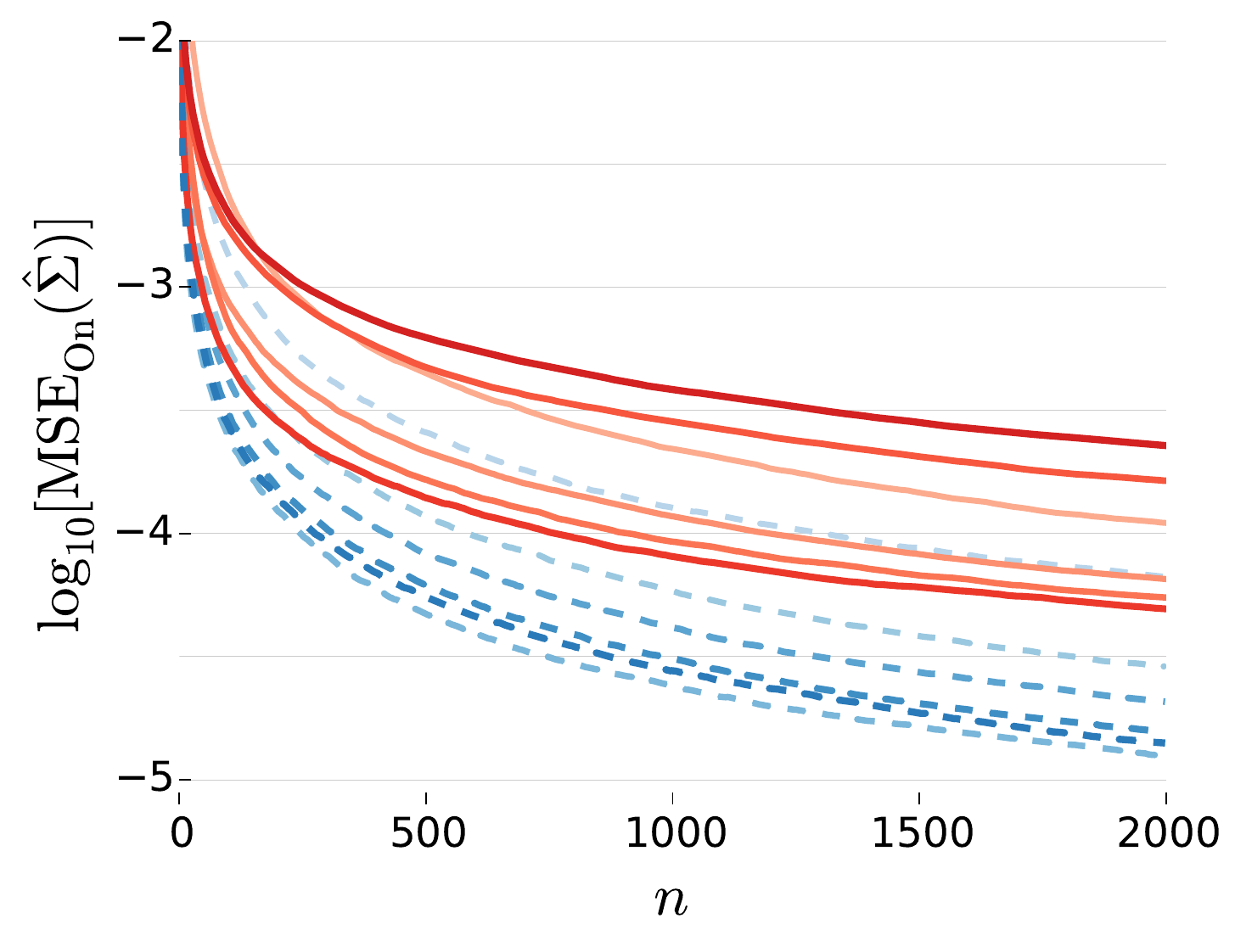} 
\includegraphics[width=5.8cm]{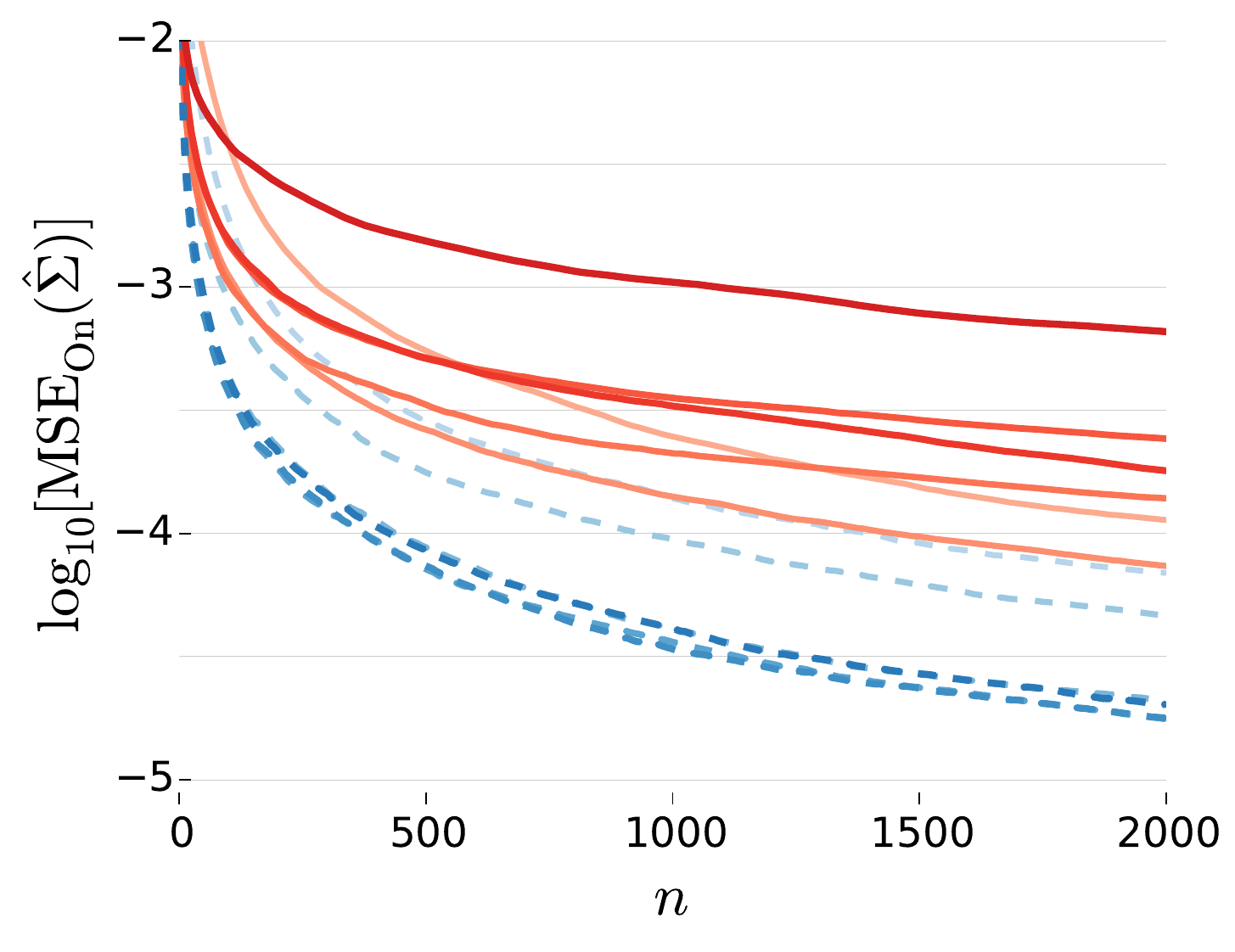} 
\includegraphics[width=5.8cm]{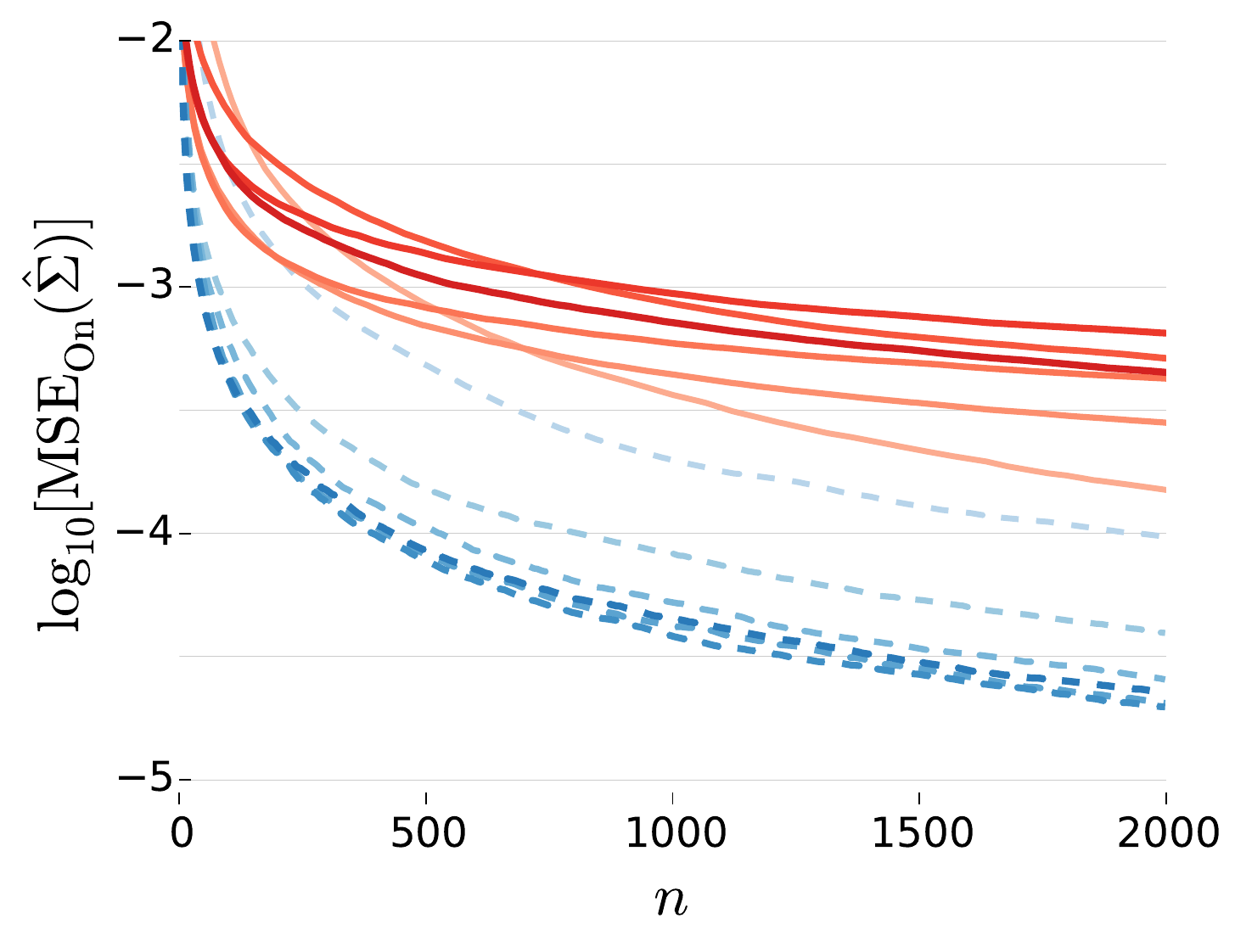} 
\caption{Covariance matrix MSE of off-diagonal elements (upper row) and on-diagonal elements (lower row), as a function of total number of samples collected. HMC is indicated by the solid orange/red lines, while CHMC is indicated by the dotted blue lines. The data is shown for various values of step size $h$, from 0.01 (light blue/light orange) to 0.25 (dark blue/ dark red). The left column is for the uniform-type, middle column is for Toeplitz-linear, and right column is for Toeplitz-geometric.}
\label{fig:beta=1}
\end{figure*}

\begin{figure}
\centering
\includegraphics[width=2.8cm]{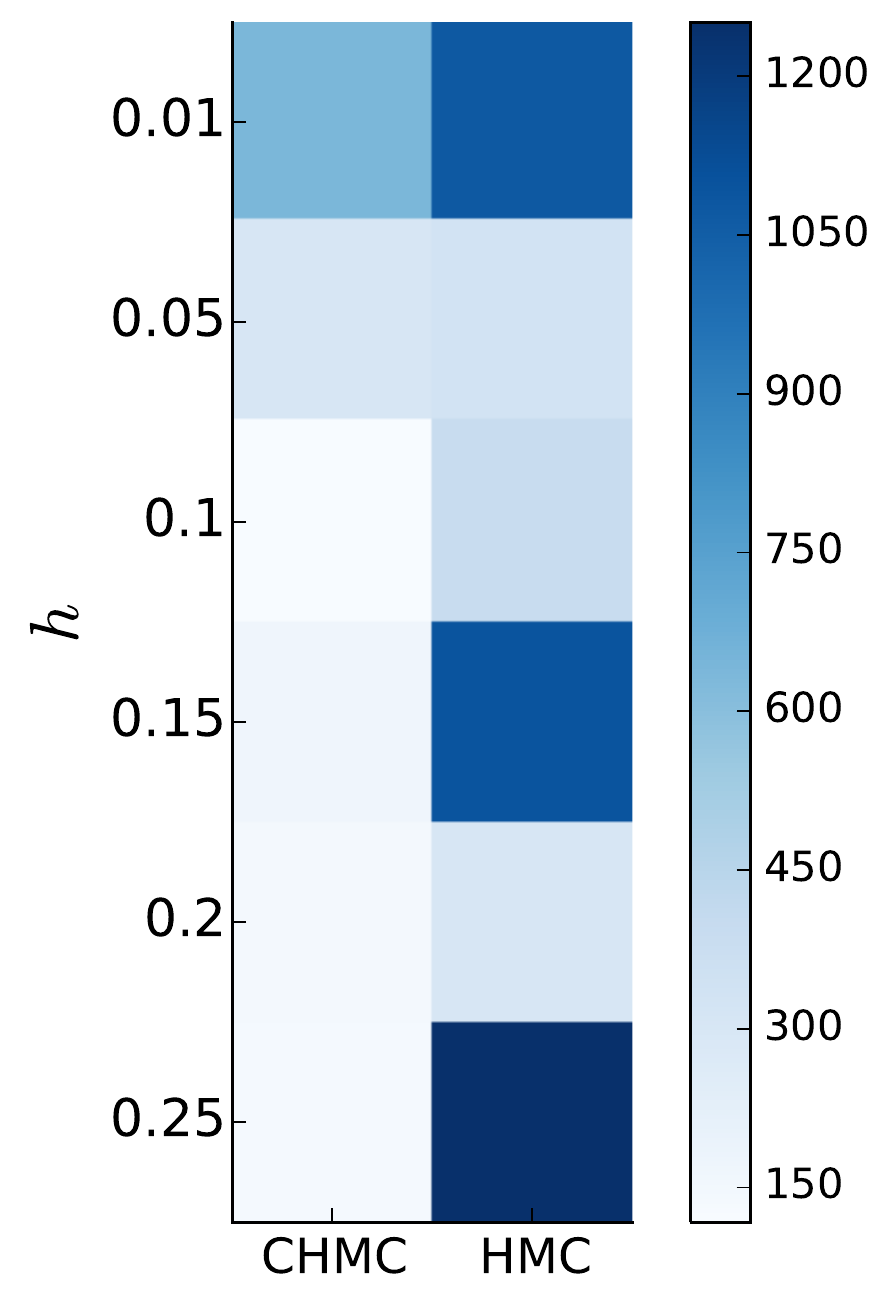}
\includegraphics[width=2.8cm]{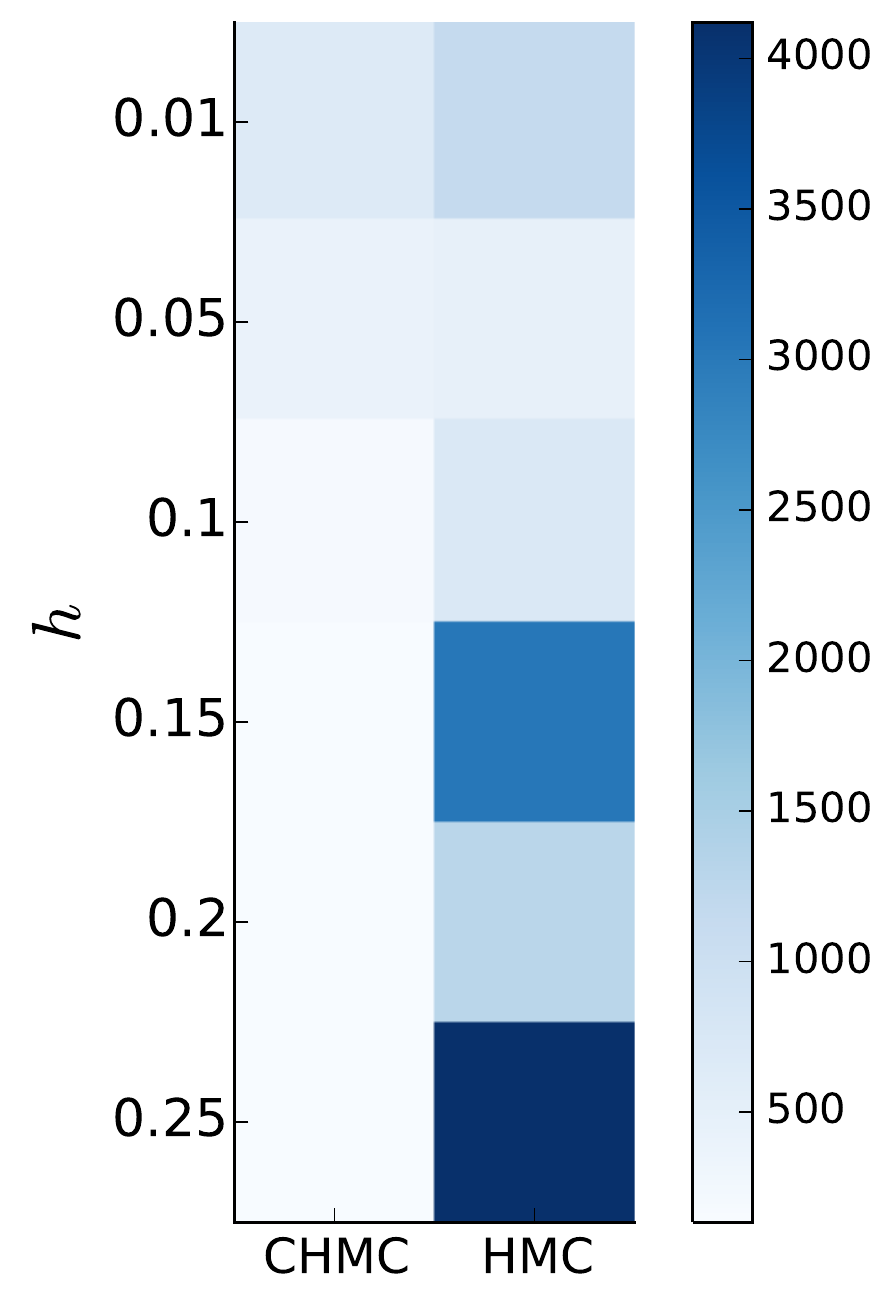}
\includegraphics[width=2.8cm]{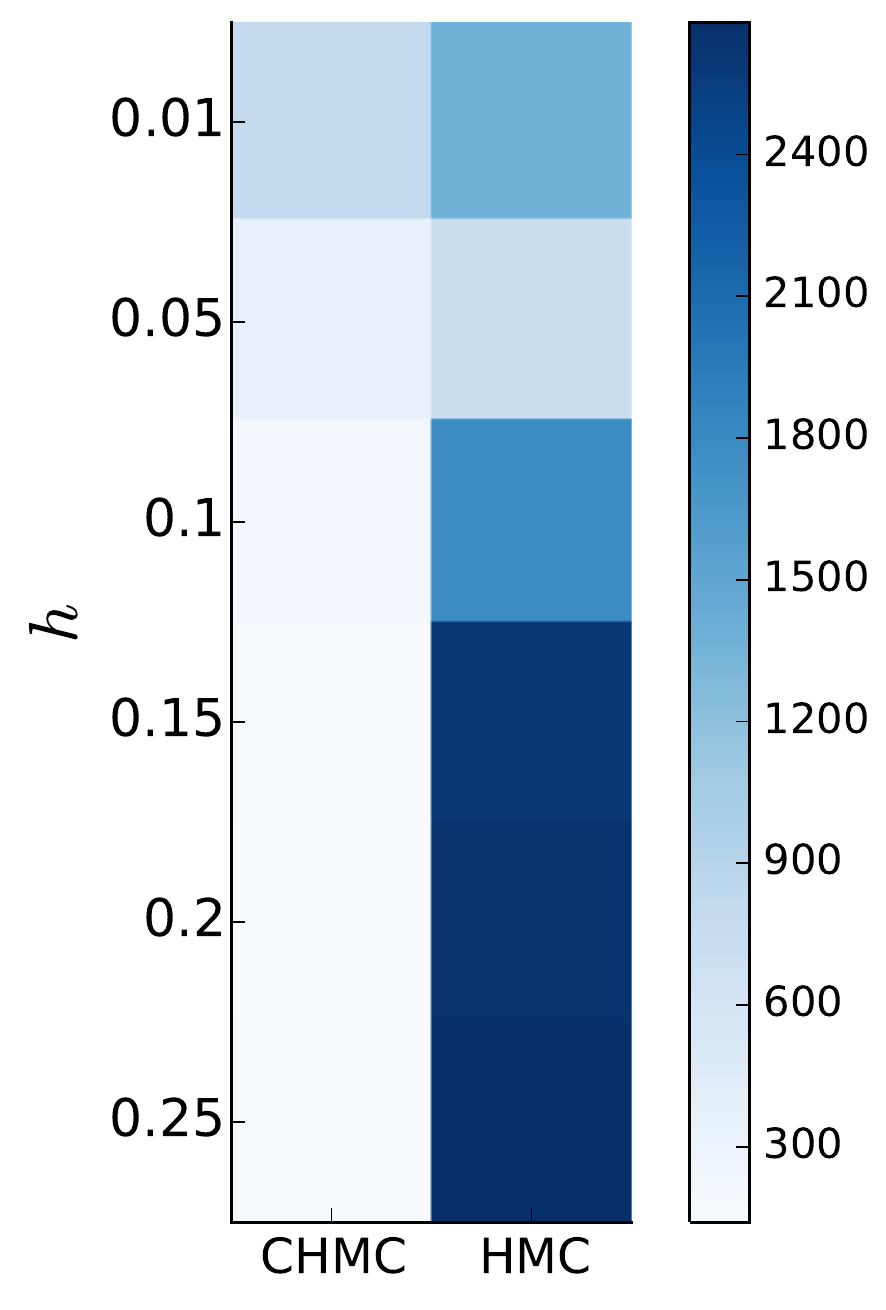}
\caption{Number of collected samples needed to reach an off-diagonal MSE $<$ 1e-4 for uniform, Toeplitz-linear, and Toeplitz-geometric correlation matrices, respectively, for all step sizes $h$.}
\label{fig:first_hits}
\end{figure}

Mean squared error data for both HMC and CHMC sampling are shown in Figure \ref{fig:beta=1}. For all three covariance structures, CHMC vastly outperforms HMC for nearly all step sizes $h$. An idea of the reduction in computation time can be ascertained by following a particular error value horizontally, noting when each of the curves intersects it. We give an indication of this in Figure \ref{fig:first_hits}, where we plot the number of samples needed for the off-diagaonal elements to reach an MSE of less than 1e-4 for different step sizes $h$. On average, a computational savings between 5x and 10x is found for CHMC over traditional HMC.

\begin{figure}[htb]
\centering
\includegraphics[width=7cm]{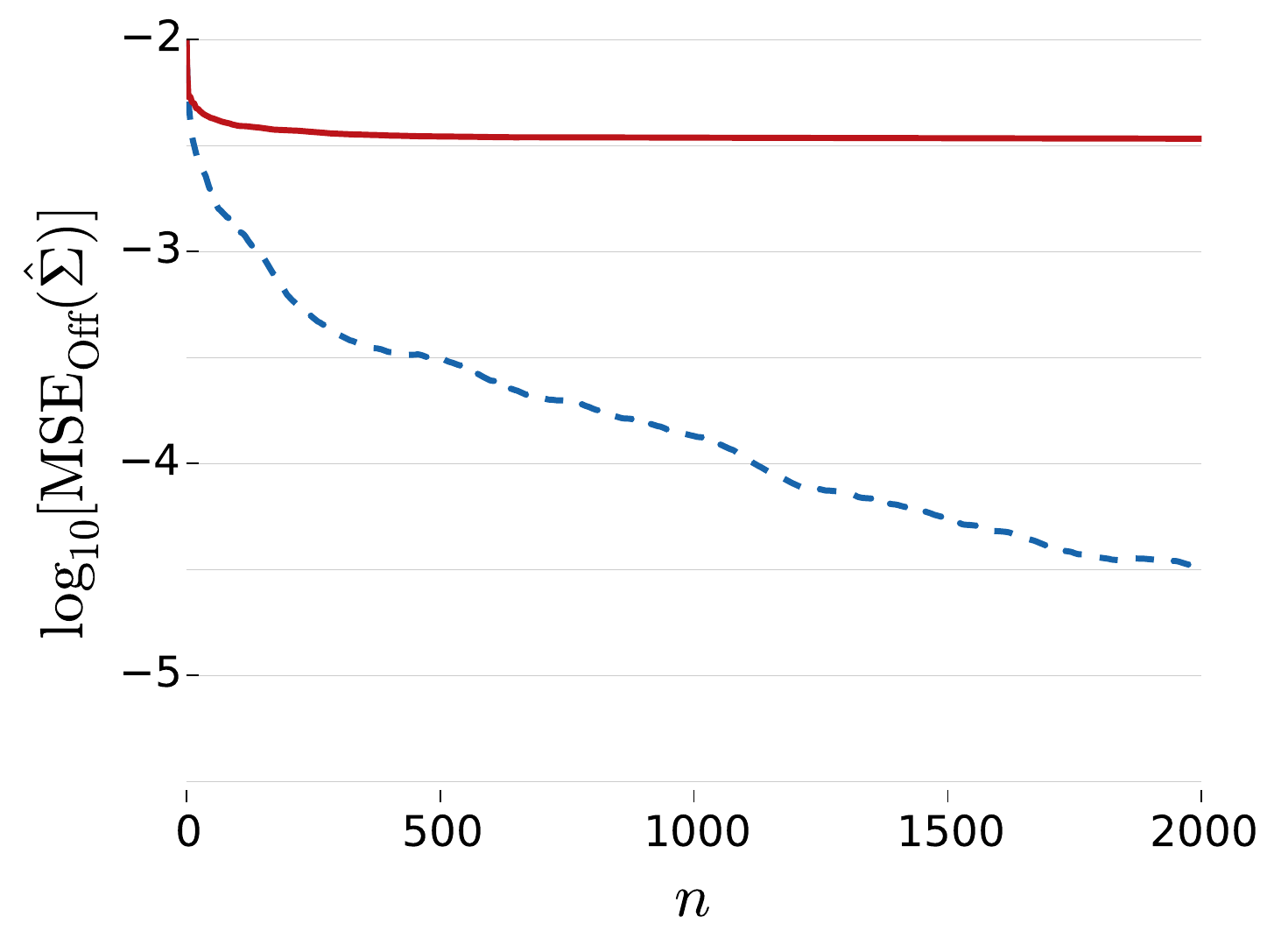}
\caption{MSE of off-diagonal elements for a particular ${\mathbf \Sigma}$ of the Toeplitz-Linear type; $h = 0.2$. HMC is shown in solid red, while CHMC is shown by the dotted blue line.}
\label{fig:bad_mixing_example}
\end{figure}

We also note that the error estimates for CHMC generally improve asymptotically for larger $h$, while for HMC, the ideal step size is problem-dependent. It is known that for traditional HMC, a poorly chosen step size can cause the sampling trajectory to get ``stuck'' in a submanifold of the target distribution-- despite $\bf P$ randomization. This often necessitates trial runs to determine the ``ideal'' step size, or to adaptively change the step size throughout the sampling procedure. The strong mixing properties provided by the ${\bf P}$-coupling in the CHMC method make the sampling routines far less susceptible to these drawbacks. We show a particular instance of a poor step size in Figure \ref{fig:bad_mixing_example}, where ${\mathbf \Sigma}$ improves steadily for CHMC as more samples are taken, but there is virtually no improvement for traditional HMC after 500 steps.

An interesting feature of CHMC is that covariance estimates improve with sample size even without momentum re-sampling. To illustrate this, we compute the MSE for Toeplitz-Linear covariance matrices in which the covariance matrices were drawn simply by Hamiltonian dynamics alone. That is, ${\bf P}_{n+1}$ is treated just as ${\bf X}_{n+1}$; it is set to ${\bf P^*}$ if $\gamma < \exp(-\Delta E)$, and ${\bf P}_n$ otherwise -- it is not re-sampled at every step from $\Xi({\bf P})$. 

\begin{figure}[htp]
\centering
\includegraphics[width=7cm]{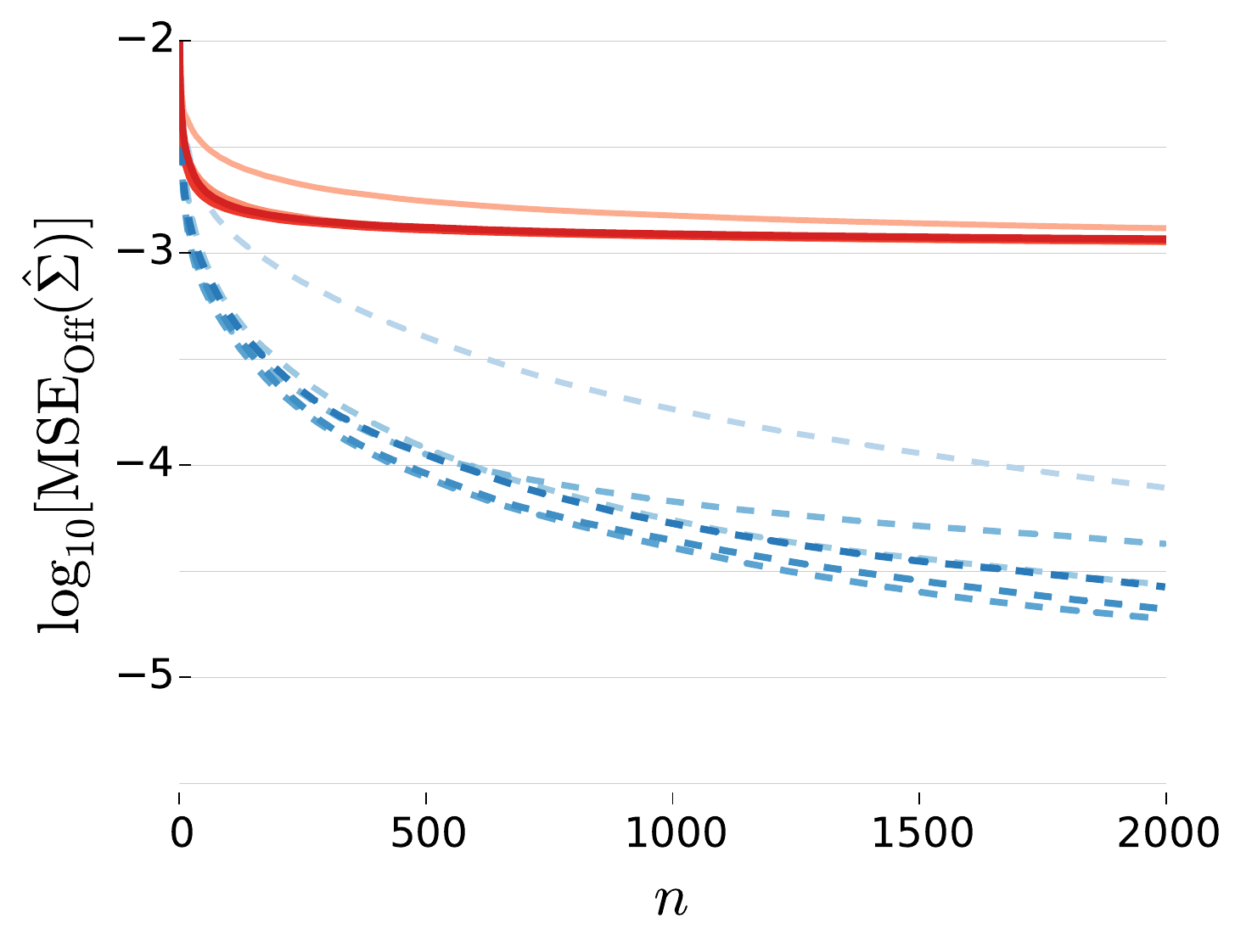} \\
\includegraphics[width=7cm]{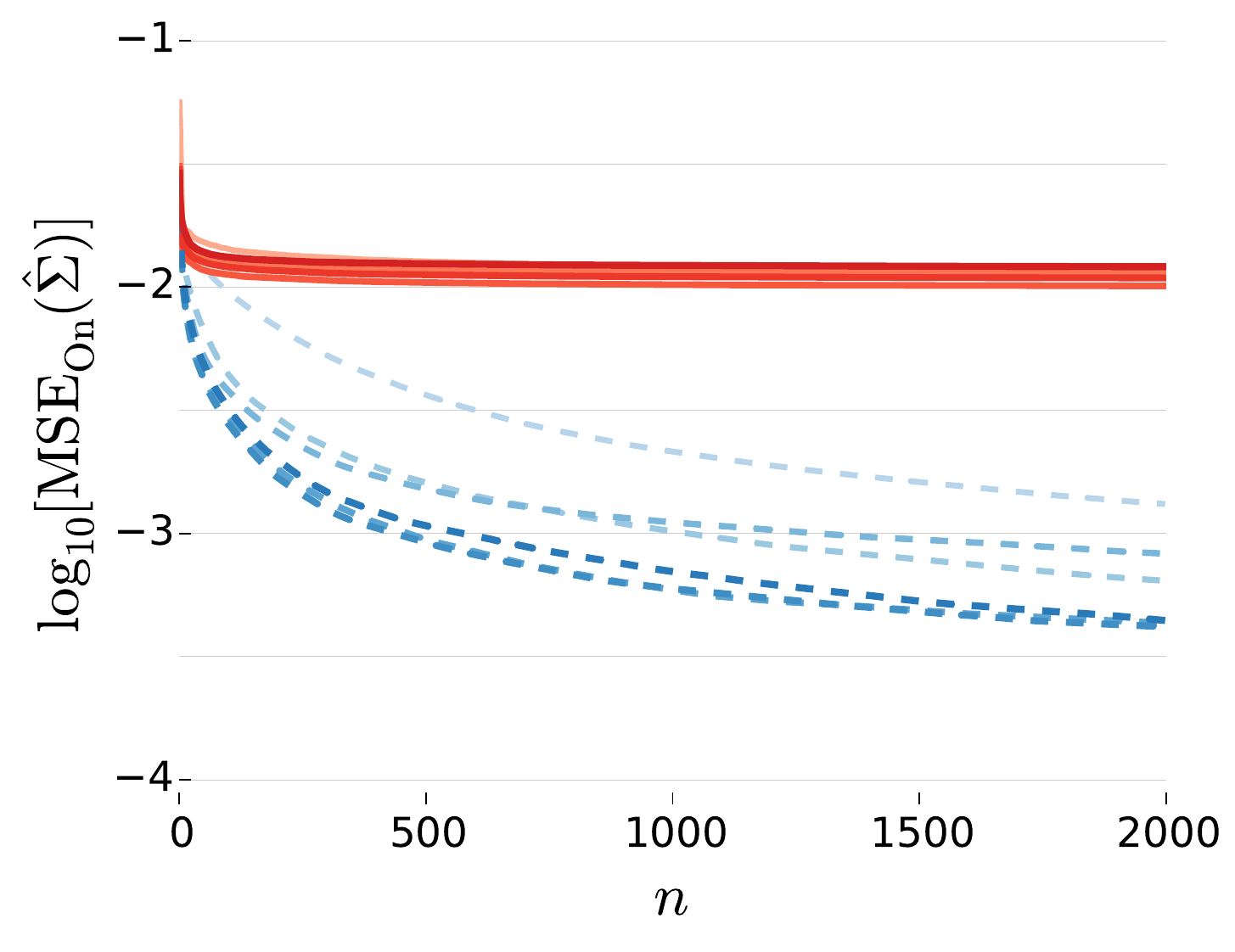} 
\caption{Toeplitz-linear correlation matrix MSE of off-diagonal elements and on-diagonal elements, respectively, versus samples collected $n$ for sampling without momentum randomization. HMC is indicated by the solid orange/red lines, while CHMC is indicated by the dotted blue lines. The data is shown for various values of step size $h$, from 0.01 (light blue/light orange) to 0.25 (dark blue/dark red).}
\label{fig:no_p_random}
\end{figure}
The covariance matrix errors via sampling without ${\bf P}$ randomization are shown in Figure \ref{fig:no_p_random}. While traditional HMC estimates show little to no improvement after about 500 samples, CHMC estimates improve steadily as more and more samples are collected. This produces, after 2000 samples, off-diagonal MSE estimates nearly 2 orders of magnitude lower. Still, comparing to Figure \ref{fig:beta=1}, it is seen that the estimates are not quite as accurate as those produced by CHMC with momentum re-sampling, so this step may still be necessary to improve accuracy. In addition, if one wishes to adapt the method for other distributions, particularly those with nonzero higher-order moments, fat tails, or for which rare samples are important, momentum sampling may still be necessary. 

\begin{figure}
\centering
\includegraphics[width=4.2cm]{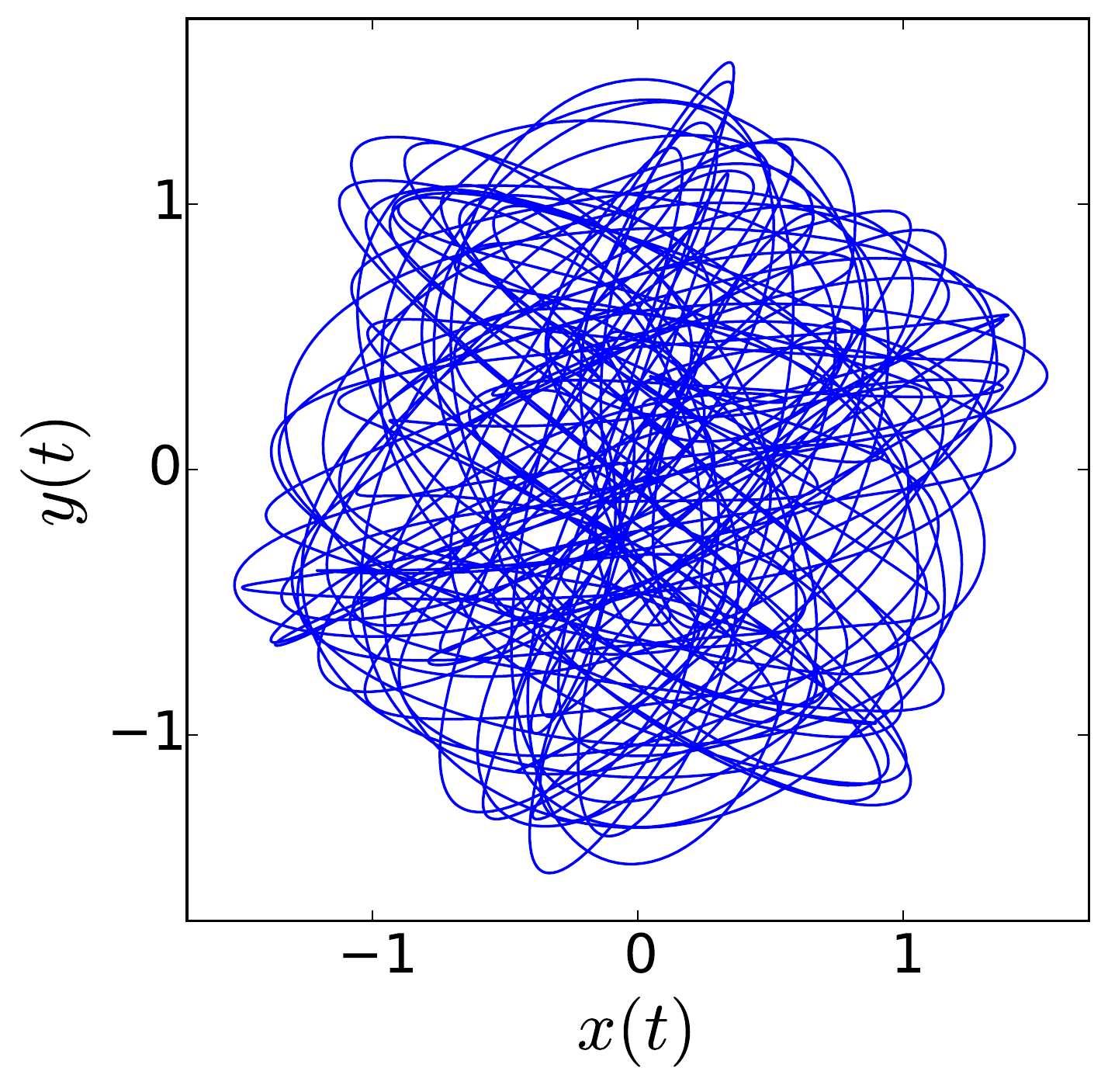}
\includegraphics[width=4.2cm]{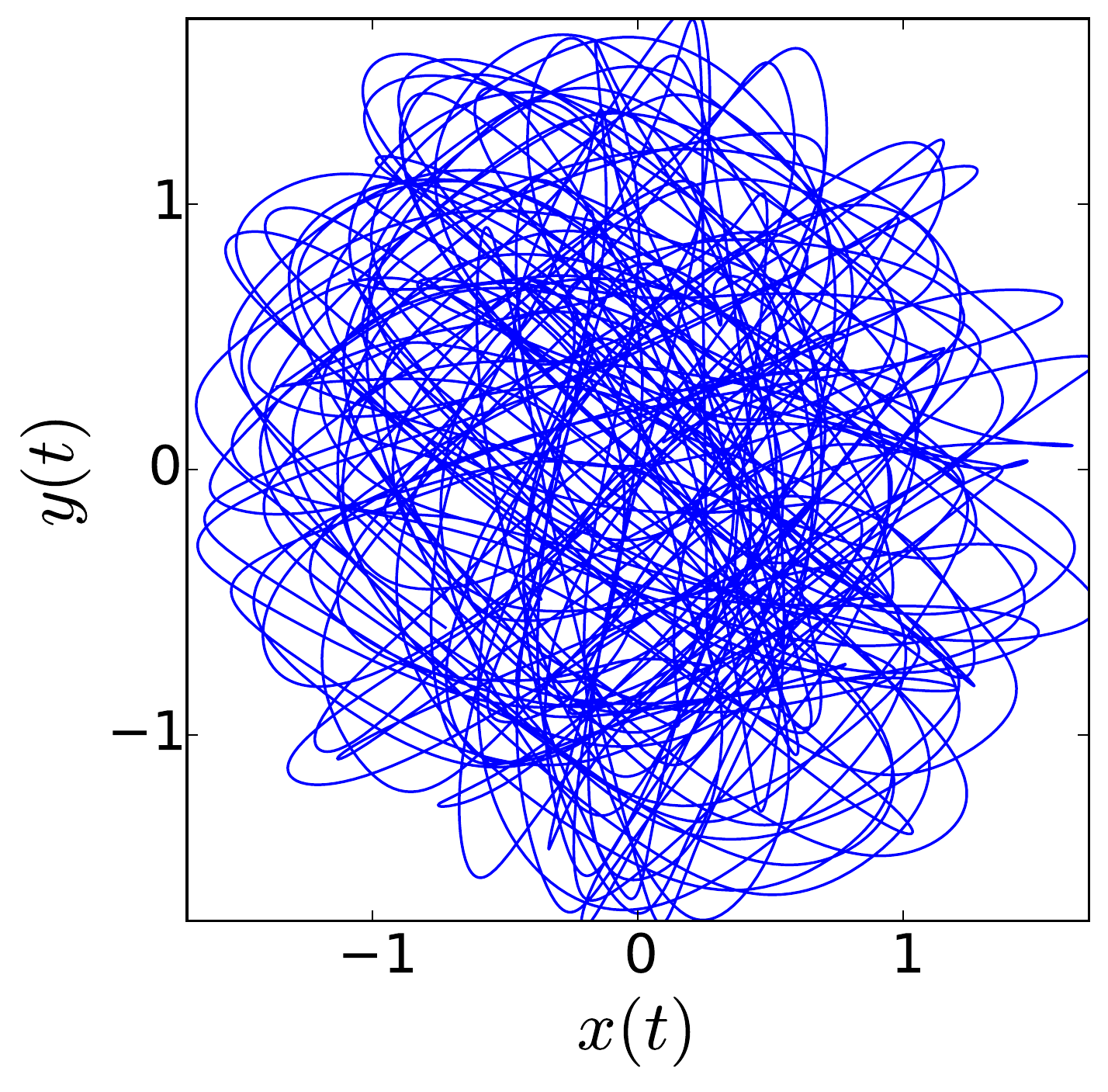} \\
\includegraphics[width=4.2cm]{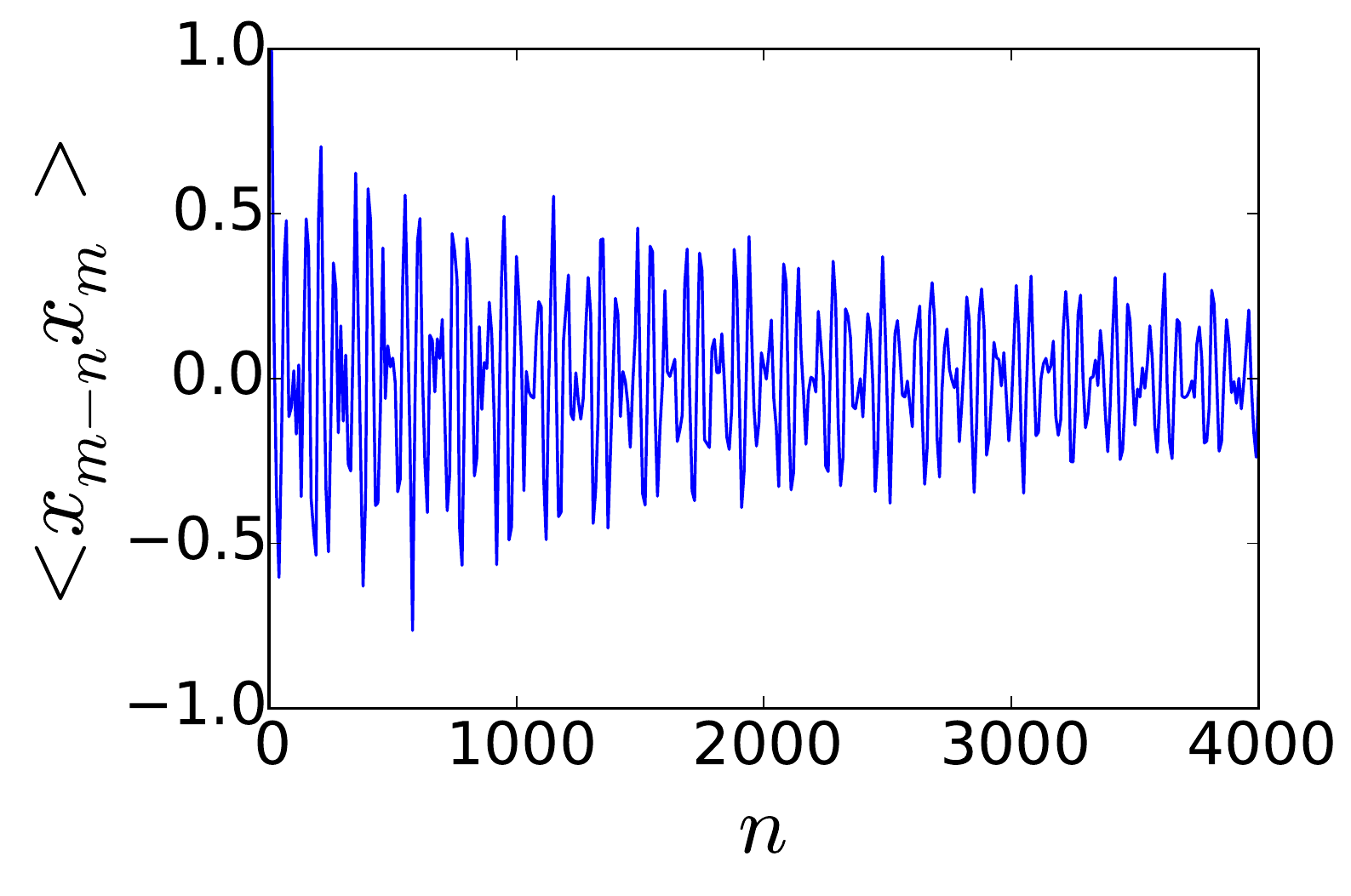} 
\includegraphics[width=4.2cm]{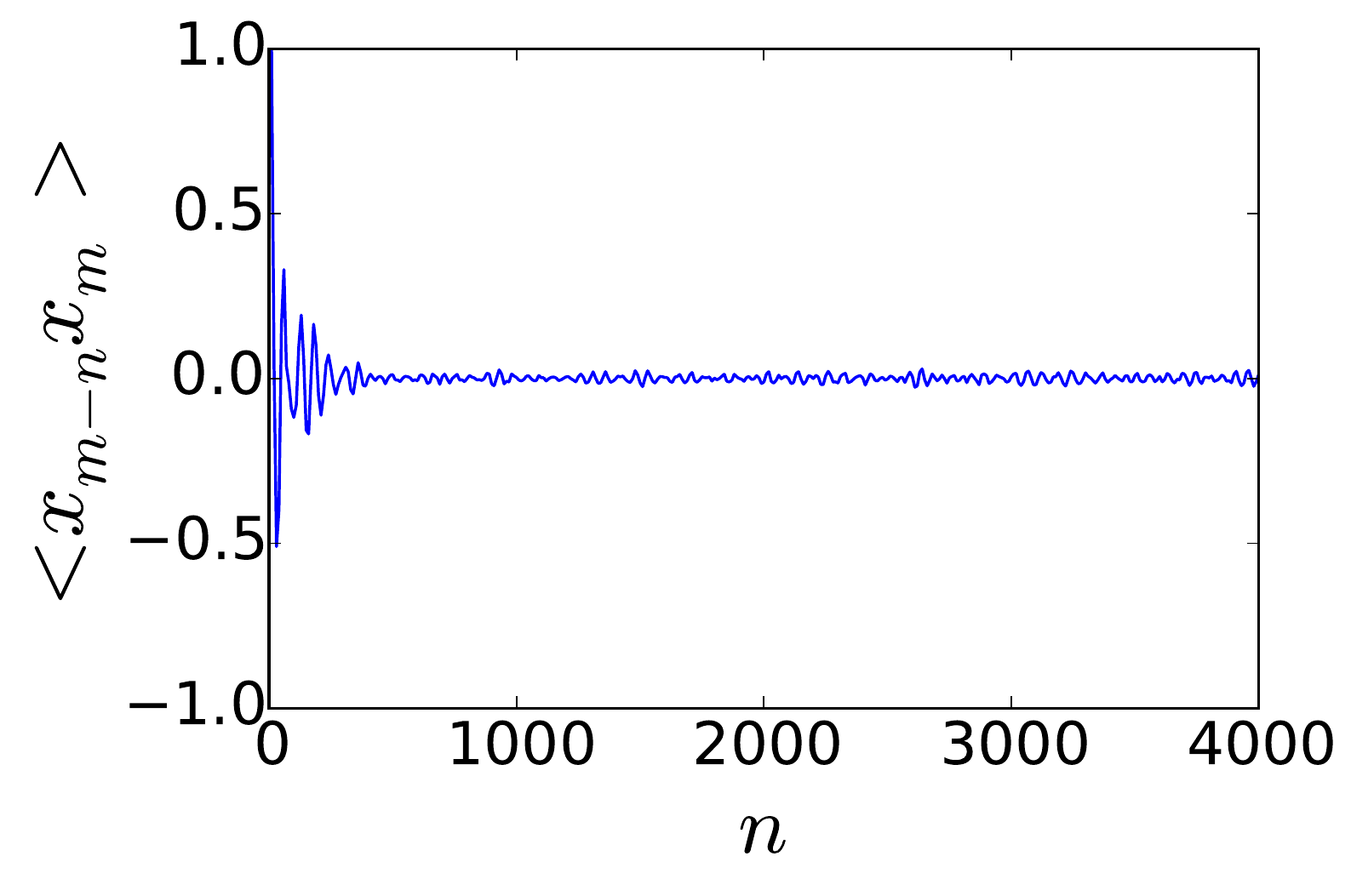} \\
\caption{{\it Top row} 2D Phase portrait projections for a (4+4)$D$ system with canonical momentum kernel (left) and chaotic momentum kernel (right). {\it Bottom row} Autocorrelations for the respective systems, averaged over all 4 position variables and over 100 different initial conditions.}
\label{fig:phase_plots_4D_uniform}
\end{figure}

The steady reduction in MSE for CHMC sampling without $\bf P$ randomization is intimately related to exponential decay of autocorrelation that is a signature of highly mixing (as opposed to simply ergodic) systems. This distinction can be belied by phase plane projections alone. This is readily seen in Figure \ref{fig:phase_plots_4D_uniform}, an example of a (4+4)$D$ system with off-diagonal correlations chosen uniformly from [-0.9,0.9]. 
The phase plane projection looks nearly as disordered for the canonical momentum kernel as for the chaotic one. Nevertheless, the correlations (here, averaged for 100 walkers in each system) decay to zero considerably faster for CHMC. This rapid decay underlies the steady improvement in accuracy for CHMC, even without momentum resampling.

\section{Summary and Future Directions}
We have shown that a non-canonical choice of momentum kernel in a hybrid Monte Carlo scheme can more effectively sample mutivariate normal distributions of various correlation structure. The kernel effectively couples momenta in pairs via quartic interaction terms, giving rise to dynamical trajectories that are chaotic and highly mixing. A key to the computational effectiveness of this momentum distribution is that it can be sampled with rejection sampling, and this method can be scaled to arbitrary dimensions without sacrificing acceptance probability. This scheme is shown to give computational savings up to ten times for various covariance matrix structures. In addition, we have shown that effective sampling can be done in the HMC scheme even without momentum resampling. 


The most obvious extension moving forward is applicability to more general distributions. On the one hand, it is not hard to produce Hamiltonian systems of appreciable dimension that exhibit chaos; on the other hand, the prevalance of such chaotic behavior is highly sytem-dependent and often hard to characterize~\cite{ott}. For example, the Hamitonian system $H = p_x^2/ + p_y^2/2 + x^2y^2/2$, closely related to the sytem in this paper and long considered an analytic candidate for globally chaotic dynamics, was eventually shown to exhibit a stable island occupying a tiny fraction of phase space~\cite{dahlqvist}. For the purposes of HMC, the goal should therefore center on finding momentum kernels that {\it reduce}, as much as possible, the prevalance and size of invariant KAM tori, opening up the available initial conditions for chaotic motion~\cite{ott}. 

Despite these difficulties, it is conceivable that general schemes could be made for other common distributions of particular functional forms. One promising extension of this method is in sampling high-dimensional integrals that arise in state and parameter estimations of geophysical and fluid models in data assimilation~\cite{abarbanelbook,ye2014precision,ye2015physrev, rey2014}. In such problems, many of the dynamical models involve variables that appear at most quadratically in the argument of a Gaussian error term, producing quartic potentials:
\begin{align*}
U({\bf X} = \{x_a(n)\}) &\sim R_f\sum_{a,n}( x_a(n+1) - f_a({\bf x}(n)))^2 \\
&+ R_m\sum_{l,n}(x_l(n) - y_l(n))^2 \\
f_a({\bf x}(n)) &\sim \sum_{b,c}C^a_{bc}x_b(n)x_c(n) + \sum_{b}D^a_{b}x_b(n)
\end{align*}
Here $f_a$ is the dynamical forward mapping and $x_a(n)$ represent the state of the system's $a$th component at discrete time $n$. The first term of the potential energy enforces model dynamics, while the second incorporates measurement errors from the data $y_l(n)$. Since the dynamics is limited to quadratic couplings for many models of interest, developing the appropriate momentum distributions for generic quartic potential energies would be of great use in these data assimilation contexts. On the other hand, the presence of cubic and quartic couplings among the position variables may be sufficient to render the HMC routine chaotic with a canonical momentum kernel alone. Further studies would be needed to characterize the Lyapunov spectrum for such systems and to investigate whether the maximal positive exponent could be tuned with more exotic momentum distributions.
 
Finally, we anticipate that the ideas presented in this paper may be of use in estimating the parameters of Bayesian posteriors that arise in artificial neural network models. These parameters typically include synaptic weights $w_{h}$ connecting nodes to one another and biases $\theta_i$ dictating response thresholds. In these models, it has proven convenient and fruitful to represent $w_{h}$ and $\theta_i$  as random variables with Gaussian distributions whose statistics are in turn controlled by hyperparameters $\tau, \tau_w, \tau_\theta$. In these cases, the potential energy function assumes the following schematic form~\cite{neal1996}:
\begin{align*}
\quad U({\bf X} = \{w_{h}, \theta_k\}) &\sim -\tau \sum_c (y_i^{(c)} - v^{O,(c)}(w_h,\theta_k))^2/2 \\
\quad &- \tau_{w}\sum_{h}w_{h}^2/2 - \tau_\theta \sum_k\theta_k^2/2 
\end{align*}
Here, $v^{O, (c)}$ indicates the network output due to the $c$th input $v^{I, (c)}$; it is a nonlinear function of the network parameters, while $y^{(c)}$ are the $c$ measurements. Further work is needed to characterize whether the Hamiltonian system with a canonical kinetic energy $\sum_i p_i^2/2$ already exhibits chaotic behavior, and if so how sensitively this behavior depends on the network architecture and strength of the hyperparameters. If a quadratic kinetic energy is found to produce dynamics with zero or small Lyapunov exponents, we expect that momentum couplings such as those investigated here to be of benefit. A deeper characterization of momenta distributions that effectively sample these target distributions opens up many promising avenues of future study. \\

\section{Acknowledgments}

The author would like to thank Henry D. I. Abarbanel and Paul Rozdeba for helpful discussions. 

\bibliographystyle{ieeetr}
\bibliography{CHMC}

\begin{thebibliography}{10}

\bibitem{duane}
S.~Duane, A.~Kennedy, B.~J. Pendleton, and D.~Roweth, ``Hybrid {M}onte
  {C}arlo,'' {\em Physics Letters B}, vol.~195, no.~2, pp.~216 -- 222, 1987.

\bibitem{neal2010}
R.~Neal, ``{MCMC} using {H}amiltonian dynamics,'' in {\em Handbook of Markov
  Chain Monte Carlo} (S.~Brooks, A.~Gelman, G.~L. Jones, and X.-L. Meng, eds.),
  ch.~5, pp.~113--162, Chapman and Hall, CRC Press, 2010.

\bibitem{metropolis}
N.~Metropolis, A.~W. Rosenbluth, M.~N. Rosenbluth, A.~H. Teller, and E.~Teller,
  ``Equation of state calculations by fast computing machines,'' {\em Journal
  of Chemical Physics}, vol.~21, no.~6, 1953.

\bibitem{hastings}
W.~Hastings, ``Monte {C}arlo samping methods using {M}arkov chains and their
  applications,'' {\em Biometrika}, vol.~57, pp.~97--109, 1970.

\bibitem{neal1996}
R.~M. Neal, {\em Bayesian Learning for Neural Networks}.
\newblock Secaucus, NJ, USA: Springer-Verlag New York, Inc., 1996.

\bibitem{mackay2003}
D.~J.~C. MacKay, {\em Information Theory, Inference, and Learning Algorithms}.
\newblock Cambridge University Press, 2003.

\bibitem{ishwaran}
H.~Ishwaran, ``Applications of hybrid {Monte Carlo} to {Bayesian} generalized
  linear models: Quasicomplete separation and neural networks,'' {\em Journal
  of Computational and Graphical Statistics}, vol.~8, no.~4, pp.~779--799,
  1999.

\bibitem{mackenze}
P.~B. Mackenze, ``An improved hybrid {M}onte {C}arlo method,'' {\em Physics
  Letters B}, vol.~226, no.~3–4, pp.~369 -- 371, 1989.

\bibitem{beskos}
A.~Beskos, N.~Pillai, G.~Roberts, J.-M. Sanz-Serna, and A.~Stuart, ``Optimal
  tuning of hybrid {M}onte {C}arlo algorithm,'' {\em Bernoulli}, vol.~19,
  pp.~1501--1534, 2013.

\bibitem{CHMC}
Y.~Fang, J.-M. Sanz-Serna, and R.~D. Skeel, ``Compressible generalized hybrid
  {M}onte {C}arlo,'' {\em Journal of Chemical Physics}, vol.~140, p.~174108,
  2014.

\bibitem{hmcdetailbalance}
J.~Sohl-Dickstein, M.~Mudigonda, and M.~Deweese, ``Hamiltonian {M}onte {C}arlo
  without detailed balance,'' in {\em Proceedings of the 31st International
  Conference on Machine Learning (ICML-14)} (T.~Jebara and E.~P. Xing, eds.),
  pp.~719--726, JMLR Workshop and Conference Proceedings, 2014.

\bibitem{shadowhmc}
J.~A. Izaguirre and S.~S. Hampton, ``Shadow hybrid {M}onte {C}arlo: An
  efficient propagator in phase space of macromolecules,'' {\em Journal of
  Computational Physics}, vol.~200, pp.~581--604, Nov. 2004.

\bibitem{exponentialhmc}
W.-L. Chao, J.~Solomon, D.~L. Michels, and F.~Sha, ``Exponential integration
  for {H}amiltonian {M}onte {C}arlo,'' in {\em International Conference on
  Machine Learning – ICML 2015}, pp.~1142--1151, July 2015.

\bibitem{highDgaussians}
F.~Orieux, O.~Feron, and J.~F. Giovannelli, ``Sampling high-dimensional
  {G}aussian distributions for general linear inverse problems,'' {\em IEEE
  Signal Processing Letters}, vol.~19, pp.~251--254, May 2012.

\bibitem{gaussian}
D.~B. Thomas, W.~Luk, P.~H. Leong, and J.~D. Villasenor, ``Gaussian random
  number generators,'' {\em ACM Computing Surveys (CSUR)}, vol.~39, no.~4,
  p.~11, 2007.

\bibitem{goldstein}
H.~Goldstein, C.~Poole, and J.~Safko, {\em Classical Mechanics}.
\newblock Addison Wesley, 2002.

\bibitem{hairer}
E.~Hairer, G.~Wanner, and C.~Lubich, {\em Geometric Numerical Integration:
  Structure-Preserving Algorithms for Ordinary Differential Equations}.
\newblock Springer Series in Computational Mathematics; Volume 31, 2nd~ed.,
  2006.

\bibitem{reichnosehoover}
B.~Leimkuhler and S.~Reich, ``A {M}etropolis adjusted {Nos{\'e}-Hoover}
  thermostat,'' {\em Mathematical Modelling and Numerical Analysis}, vol.~43,
  pp.~743--755, 2009.

\bibitem{ergodichierarchy}
J.~Berkovitz, R.~Frigg, and F.~Kronz, ``The ergodic hierarchy, randomness and
  {H}amiltonian chaos,'' {\em Studies in History and Philosophy of Science Part
  B}, vol.~37, no.~4, pp.~661--691, 2006.

\bibitem{mixingbook}
R.~Sturman, J.~M. Ottino, and S.~Wiggins, {\em The Mathematical Foundations of
  Mixing}.
\newblock Cambridge University Press, 2006.

\bibitem{ott}
E.~Ott, {\em Chaos in Dynamical Systems}.
\newblock Cambridge, New York: Cambridge University Press, 2002.

\bibitem{baladi2000}
V.~Baladi, {\em Positive Transfer Operators and Decay of Correlations}, vol.~16
  of {\em Advanced Series in Nonlinear Dynamics}.
\newblock World Scientific, 2000.

\bibitem{elegantchaos}
J.~Sprott, {\em Elegant Chaos: Algebraically Simple Chaotic Flows}.
\newblock World Scientific, 2010.

\bibitem{nonlineartimeseries}
H.~Kantz and T.~Schreiber, {\em Nonlinear Time Series Analysis}.
\newblock Cambridge Nonlinear Science Series, Cambridge, New York: Cambridge
  University Press, 1997.
\newblock Originally published: 1997.

\bibitem{montecarlomethods}
C.~P. Robert and G.~Casella, {\em Monte {C}arlo Statistical Methods (Springer
  Texts in Statistics)}.
\newblock Secaucus, NJ, USA: Springer-Verlag New York, Inc., 2005.

\bibitem{dahlqvist}
P.~Dahlqvist and G.~Russberg, ``Existence of stable orbits in the
  ${\mathit{x}}^{2}$${\mathit{y}}^{2}$ potential,'' {\em Physical Review
  Letters}, vol.~65, pp.~2837--2838, Dec 1990.

\bibitem{abarbanelbook}
H.~D. Abarbanel, {\em Predicting the Future: Completing Models of Observed
  Complex Systems}.
\newblock Springer, 2013.

\bibitem{ye2014precision}
J.~Ye, N.~Kadakia, P.~J. Rozdeba, H.~D.~I. Abarbanel, and J.~C. Quinn,
  ``Improved variational methods in statistical data assimilation,'' {\em
  Nonlinear Processes in Geophysics}, vol.~22, no.~2, pp.~205--213, 2015.

\bibitem{ye2015physrev}
J.~Ye, D.~Rey, N.~Kadakia, M.~Eldridge, U.~Morone, P.~Rozdeba, H.~D.~I.
  Abarbanel, and J.~C. Quinn, ``A systematic variational method for statistical
  nonlinear state and parameter estimation,'' {\em Physical Review E}, vol.~92,
  no.~5, p.~052901, 2015.

\bibitem{rey2014}
D.~Rey, M.~Eldridge, M.~Kostuk, H.~D. Abarbanel, J.~Schumann-Bischoff, and
  U.~Parlitz, ``Accurate state and parameter estimation in nonlinear systems
  with sparse observations,'' {\em Physics Letters A}, vol.~378, no.~11,
  pp.~869--873, 2014.

\end{thebibliography}
\end{document}